\documentclass{kybernetika}

\usepackage{amsmath}
\usepackage{amsfonts}
\usepackage{amssymb}
\usepackage{graphicx}
\usepackage{comment}
\usepackage{bm}
\usepackage{color} 

\newtheorem{theorem}{Theorem}[section]

\newtheorem{remark}[theorem]{Remark}

\newtheorem{example}[theorem]{Example}
\newtheorem{definition}[theorem]{Definition}

\usepackage{mathtools}

\def\bmtheta{{\bm{\theta}}}
\def\bmSigma{{\bm{\Sigma}}}
\def\bmmu{{\bm{\mu}}}

\newcommand{\R}{\mathbb{R}}

\newcommand{\transp}{{T}}
\renewcommand{\d}{\hbox{d}}

\usepackage{graphicx}

\begin{document}
\pagestyle{myheadings}

\title{Expected Utility Maximization and Conditional Value-at-Risk Deviation-based Sharpe Ratio in Dynamic Stochastic Portfolio Optimization}

\author{So\v{n}a Kilianov\'a and Daniel \v{S}ev\v{c}ovi\v{c}}
\contact{So\v{n}a}{Kilianov\'a}{Department of Applied Mathematics and Statistics, FMFI, Comenius University, Mlynsk\'a dolina, 84248 Bratislava, Slovakia}{kilianova@fmph.uniba.sk}
\contact{Daniel}{\v{S}ev\v{c}ovi\v{c}}{Department of Applied Mathematics and Statistics, FMFI, Comenius University, Mlynsk\'a dolina, 84248 Bratislava, Slovakia}{sevcovic@fmph.uniba.sk}

\markboth{S. Kilianov\'a and D. \v{S}ev\v{c}ovi\v{c}} {Expected utility maximization and conditional value-at-risk deviation}

\maketitle

\begin{abstract}
In this paper we investigate the expected terminal utility maximization approach for a dynamic stochastic portfolio optimization problem. We solve it numerically by solving an evolutionary Hamilton-Jacobi-Bellman equation which is transformed by means of the Riccati transformation. We examine the dependence of the results on the shape of a chosen utility function in regard to the associated risk aversion level. We define the  
 Conditional value-at-risk deviation ($CVaRD$) based Sharpe ratio for measuring risk-adjusted performance of a dynamic portfolio. We compute optimal strategies for a portfolio investment problem motivated by the German DAX 30 Index and we evaluate and analyze the dependence of the $CVaRD$-based Sharpe ratio on the utility function and the associated risk aversion level.

\end{abstract}



\section{Introduction}

Expected utility theory in relation with risk measures are an interesting topic and have been investigated by many authors in the literature. It is useful for investors in practice to know relationships between the maximal utility approach and risk measures like for example value-at-risk ($VaR$) or conditional value-at-risk ($CVaR$). Seck et al. in \cite{Seck, AndrieuLaraSeck} examine connection between risk measures and parameterized families of loss aversion utility functions. For each risk measure they provide a class of associated utility functions, for which the problem of maximizing profit subject to a risk measure constraint and the max-min problem of optimization over the class of utility functions are equivalent. They find parameters of the considered utility functions which are equivalent to a given level of corresponding risk measure. In Zheng \cite{Zheng}, the efficient frontier problem of maximizing the expected utility of terminal wealth and minimizing the $CVaR$ of the utility loss has been studied. The authors look for the optimal trade-off between the utility
maximization and the $CVaR$ minimization by penalizing one or the other via a weighting parameter. A relationship between risk measures and the utility concept is studied by Denuit et al. in \cite{Denuit} and other sources as well. 

In this paper, we investigate the impact of non-constant risk aversion on the outcome of multi-period portfolio optimization. We do so by evaluating an alternative Sharpe ratio with the $CVaRD$ measure in the denominator instead of standard deviation. To our knowledge, the proposed $CVARD$-based Sharpe ratio is a novel concept and so far it has not been related to the utility maximization approach. We analyze how the $CVaRD$-based Sharpe ratio depends on the shape of a chosen utility function with focus on the associated risk aversion parameter. The method utilized for the expected utility maximization problem is based on solving the evolutionary Hamilton-Jacobi-Bellman equation numerically, which is first transformed by means of the Riccati transformation. We compute optimal strategies for a portfolio investment problem motivated by the German DAX 30 Index and we evaluate and analyze the dependence of the $CVaRD$-based Sharpe ratio on the utility function and risk aversion. 

As a tool for solving the associated terminal utility maximization problems, we ge\-ne\-ra\-li\-ze the numerical method from Kilianov\'a and \v{S}ev\v{c}ovi\v{c} \cite{KilianovaSevcovicANZIAM}, where the authors have considered a dynamic portfolio selection problem with regular savings enabled. We proposed and analyzed solution to the corresponding Hamilton-Jacobi-Bellman (HJB) equation by utilizing a Riccati transformation of the value function of the optimization problem. In this paper, we generalize the transformation process to more general processes with arbitrary drift and volatility function. This allows for considering more general processes like e.g. standard dynamic portfolio selection problem studied in the paper mentioned above or worst-case portfolio optimization studied in paper \cite{KilianovaTrnovska} by Ki\-lia\-no\-v\'a and Trnovsk\'a, or any other problem with arbitrary drift and volatility function. 

The paper is organized as follows. In the next section we summarize basic assumptions made on the underlying stochastic process. In section 3 we present stochastic dynamic portfolio optimization approach based on maximization of the terminal utility function. It leads to a solution to the fully nonlinear Hamilton-Jacobi-Bellman equation. In section 4 we present a method of transformation of the HJB equation into a quasi-linear parabolic equation by means of the Riccati transformation. Section 5 is devoted to analysis of the value function arising in portfolio optimization process. In section 6 we recall a numerical method for solving HJB equations. Finally, section 7 is focused on risk measures and their application in evaluation of portfolio performance. We introduce the so-called $CVaRD$-based Sharpe ratio and we calculate this ratio for optimal portfolio selection based on utility maximization approach, for various risk aversion setups.

\section{Generalized stochastic process} 

We consider an underlying stochastic process $\{x_t\}$ satisfying the stochastic differential equation (SDE):
\begin{equation}
\label{process_x}
d x_t = \mu(x_t, t, \bmtheta_t) dt + \sigma(x_t, t, \bmtheta_t) dW_t
\end{equation}
where the control process $\{\bmtheta_t\}$ is adopted to the process $\{x_t\}$. Here $\{W_t\}$ is the standard Wiener process and functions $(x,t,\bmtheta)\mapsto\mu(x,t,\bmtheta)$ and $(x,t,\bmtheta )\mapsto\sigma(x,t,\bmtheta)^2$ are $C^1$ smooth in $x, t$ and $\bmtheta$ variables.

\begin{example}
As an example of the stochastic process (\ref{process_x}) one can consider a portfolio optimization problem with regular saving. In this case, the volatility function is given by
$\sigma(x,t,\bmtheta)^2 =\bmtheta^T \bmSigma  \bmtheta$, 
where $\bmSigma$ is a positive definite covariance matrix of asset returns. The drift function is given by $\mu(x,t,\bmtheta) = \bmmu^T\bmtheta - \frac12 \sigma(x,t,\bmtheta)^2  +\varepsilon e^{-x} +r$, 
where $\bmmu$ is the vector of mean returns of assets, $\varepsilon$ represents an inflow/outflow to the portfolio and it may depend on $x$ and $t$ variables. The parameter $r\ge0$ is an interest rate of a risk-less bond. The stochastic process $\{x^\bmtheta_t\}$ is a logarithmic transformation of a stochastic process $\{y_t^{\tilde\bmtheta} \}_{t\ge0}$ driven by the stochastic differential equation
$$
d y_t^{\tilde\bmtheta} = (\varepsilon + (r +\mu({\tilde\bmtheta})) y_t^{\tilde\bmtheta})
d t + \sigma(\tilde\bmtheta) y_t^{\tilde\bmtheta} d W_t, \label{processYeps}
$$
where $\tilde\bmtheta(y,t) = \bmtheta(x,t)$ with $x=\ln y$  (c.~f. Kilianov\'a and {\v S}ev{\v c}ovi{\v c} \cite{KilianovaSevcovicANZIAM}). 
\end{example}
\begin{example}
Another example arises in the so-called worst case portfolio optimization problem investigated by Kilianov\'a and Trnovsk\'a in \cite{KilianovaTrnovska} in which the volatility function is given by $\sigma(x,t,\bmtheta)^2 = \max_{\bmSigma\in{\mathcal K}}\bmtheta^T \bmSigma  \bmtheta$, 
where $\mathcal K$ is an uncertainty set of positive definite covariance matrices. The drift function is given by $\mu(x,t,\bmtheta) = \min_{\bmmu\in{\mathcal E}}\bmmu^T\bmtheta - \frac12 \sigma(x,t,\bmtheta)^2  +\varepsilon e^{-x} +r$, 
where $\mathcal E$ is a given uncertainty set of mean returns. 
\end{example}

\section{Dynamic stochastic optimization problem}

In a dynamic stochastic optimization problem, our purpose is to maximize the conditional expected value of the terminal utility of the portfolio:
\begin{equation}
\max_{\bmtheta|_{[0,T)}} \mathbb{E}
\left[U(x_T^\bmtheta)\, \big| \, x_0^\bmtheta=x_0 \right],
\label{maxproblem}
\end{equation}
where $\{x_t^{\bmtheta}\}$ is It\=o's stochastic process of the form (\ref{process_x}) on a finite time horizon $[0,T]$, $U: \mathbb{R} \to \mathbb{R}$ is a given terminal utility function and $x_0$ a given initial state condition of  $\{x_t^\bmtheta\}$ at $t=0$. The function $\bmtheta:   \mathbb{R} \times [0,T) \to \R^n$  represents an unknown control function governing the underlying stochastic process $\{x_t^\bmtheta\}$. 

We assume that the control parameter $\bmtheta$ belongs to a closed convex  subset $\Delta$ of the compact convex simplex $\mathcal{S}^n = \{\bmtheta \in \mathbb{R}^n\  |\  \bmtheta \ge \mathbf{0}, \mathbf{1}^\transp \bmtheta = 1\} \subset \mathbb{R}^n$, where $\mathbf{1} = (1,\cdots,1)^\transp \in \mathbb{R}^n$. If we introduce the value function
\begin{equation}
V(x,t):= \sup_{  \bmtheta|_{[t,T)}} 
\mathbb{E}\left[U(x_T^\bmtheta) | x_t^\bmtheta=x \right]
\end{equation}
then $V(x,T):=U(x)$. Following Bertsekas \cite{Bertsekas},  the value function $V=V(x,t)$ satisfies the fully nonlinear Hamilton-Jacobi-Bellman HJB parabolic equation (see also Kilianov\'a and {\v S}ev{\v c}ovi{\v c} \cite{KilianovaSevcovicANZIAM}): 
\begin{eqnarray}
&& \partial_t V + \max_{ \bmtheta \in \Delta} 
\left(
\mu(x,t,\bmtheta)\, \partial_x V 
+ \frac{1}{2} \sigma(x,t,\bmtheta)^2\, \partial_x^2 V \right) = 0\,, \  (x,t)\in\R\times [0,T), \label{eq_HJB}
\\
&& V(x,T)=U(x), \quad x\in\R. \label{init_eq_HJB}
\end{eqnarray}

\section{The Riccati transformation of the HJB equation to a quasi-linear equation}
\label{sec:HJB}

In the context of solving the HJB equation, the Riccati transformation was proposed by Abe and Ishimura in \cite{AI} and later studied by 
Ishimura and {\v S}ev{\v c}ovi{\v c} \cite{IshSev}, Xia \cite{Xia}, Macov\'a and \v{S}ev\v{c}ovi\v{c} \cite{MS}, Kilianov\'a and {\v S}ev{\v c}ovi{\v c} \cite{KilianovaSevcovicANZIAM}, Kilianov\'a and Trnovsk\'a \cite{KilianovaTrnovska}. The Riccati transformation of the value function $V$ is defined as follows:

\begin{equation}
\varphi(x,t) = - \frac{\partial_x^2 V(x,t)}{\partial_x V(x,t)}.
\label{eq_varphi}
\end{equation}

Suppose for a moment that the value function $V(x,t)$ is increasing in the $x$-variable. This is a natural assumption in the case when the terminal utility function $U(x)$ is an increasing function in the $x$ variable. The HJB equation (\ref{eq_HJB}) can then be rewritten as follows:
\begin{equation}
\partial_t V - \alpha(x,t,\varphi) \partial_x V = 0, \qquad V(x,T)=U(x),
\label{eq_HJBtransf}
\end{equation}
where $\alpha(x,t,\varphi)$ is the value function of the following parametric optimization problem:
\begin{equation}
\alpha(x,t,\varphi) = \min_{ \bmtheta \in \Delta} 
\left(
-\mu(x,t,\bmtheta) +  \frac{\varphi}{2}\sigma(x,t,\bmtheta)^2\right)\,.
\label{eq_alpha_def}
\end{equation}

\begin{example}\label{exmp:portfolio1}
In the stochastic dynamic portfolio optimization problem we have 
\[
\mu(x,t,\bmtheta) = \bmmu^T\bmtheta - \frac12 \bmtheta^T \bmSigma  \bmtheta  +\varepsilon e^{-x} +r, \quad\hbox{and} \quad 
\sigma(x,t,\bmtheta)^2 =\bmtheta^T \bmSigma  \bmtheta,
\]
where $\bmSigma$ is a positive definite covariance matrix. Hence the function $\alpha(x,t,\varphi)$ can be rewritten as follows: $\alpha(x,t,\varphi) = \tilde\alpha(\varphi) -\varepsilon e^{-x} -r$,
where $\tilde\alpha$ is the value function of the parametric quadratic optimization problem
\begin{equation}
\tilde\alpha(\varphi) = \min_{ \bmtheta \in \Delta} 
\left(-\bmmu^T\bmtheta +  \frac{\varphi+1}{2} \bmtheta^T \bmSigma  \bmtheta \right)\,.
\label{eq_alpha_def_quadratic}
\end{equation}
A graphical example of the function $\alpha$ in which $\bmmu$ and $\bmSigma$ were obtained from DAX30 data set is depicted in Figure \ref{fig:alpha_alphader_alphaderder}.
\begin{figure}
    \centering
    \includegraphics[width=0.35\textwidth]{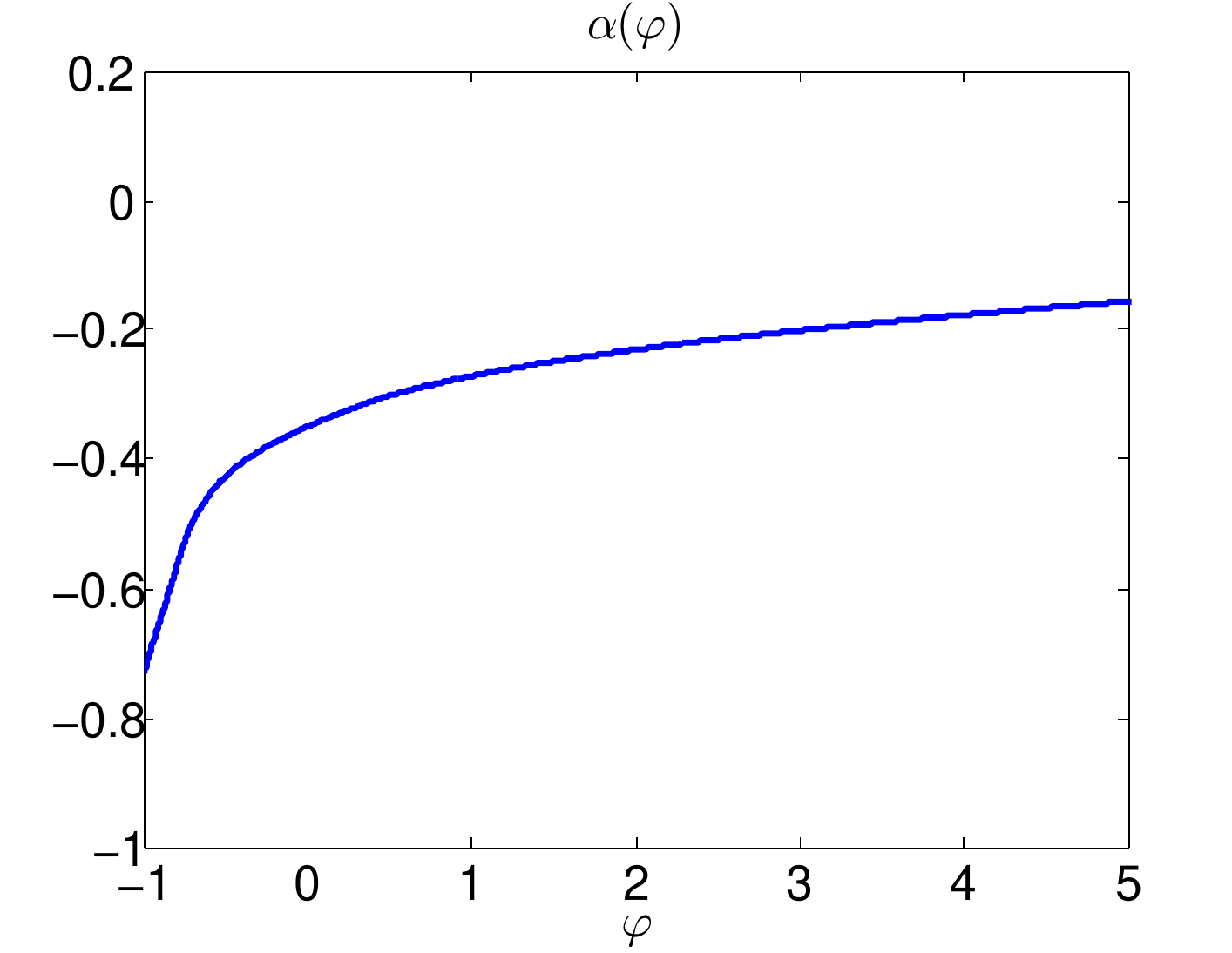}
    \hglue-0.5truecm
    \includegraphics[width=0.35\textwidth]{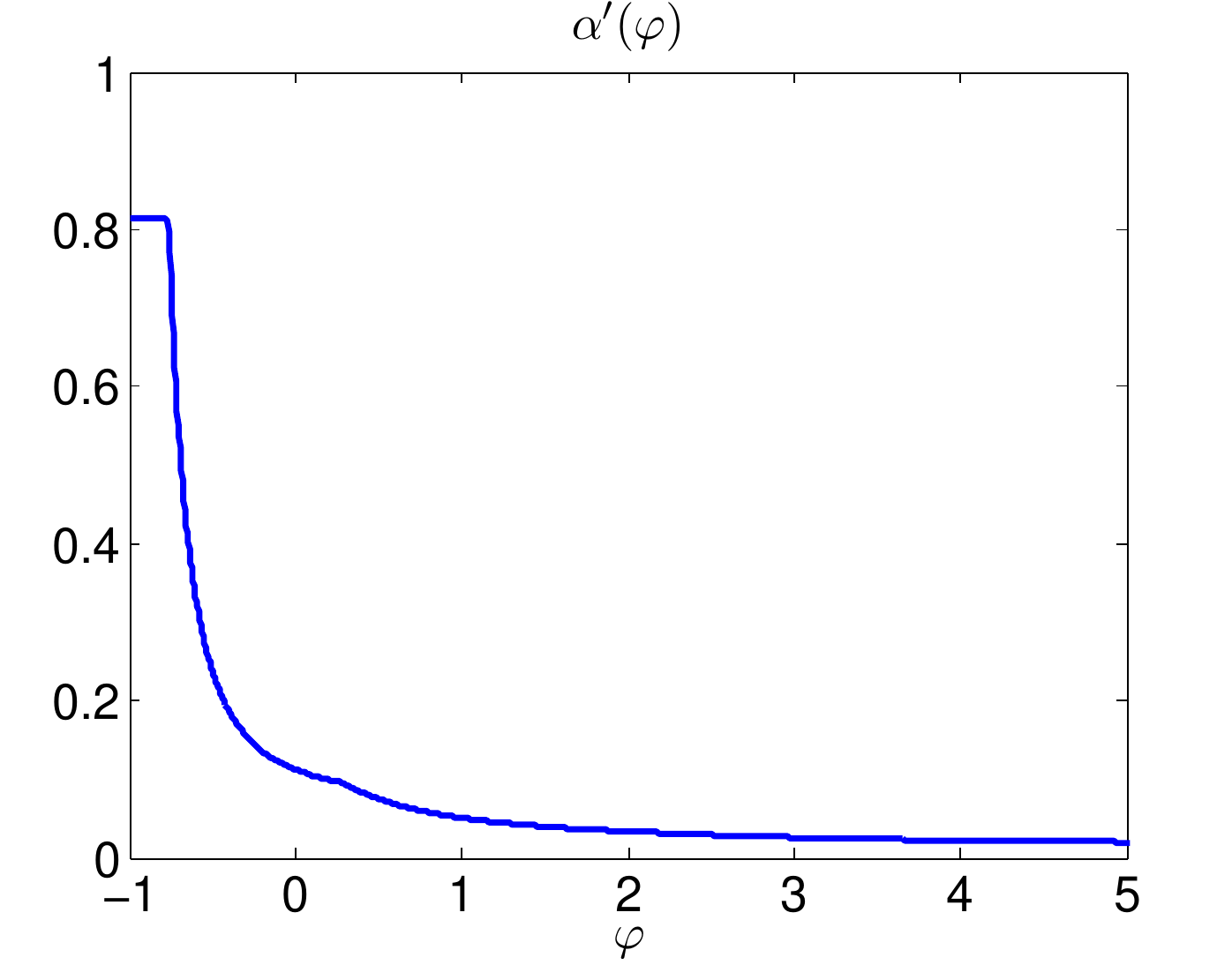}
    \hglue-0.5truecm
    \includegraphics[width=0.34\textwidth]{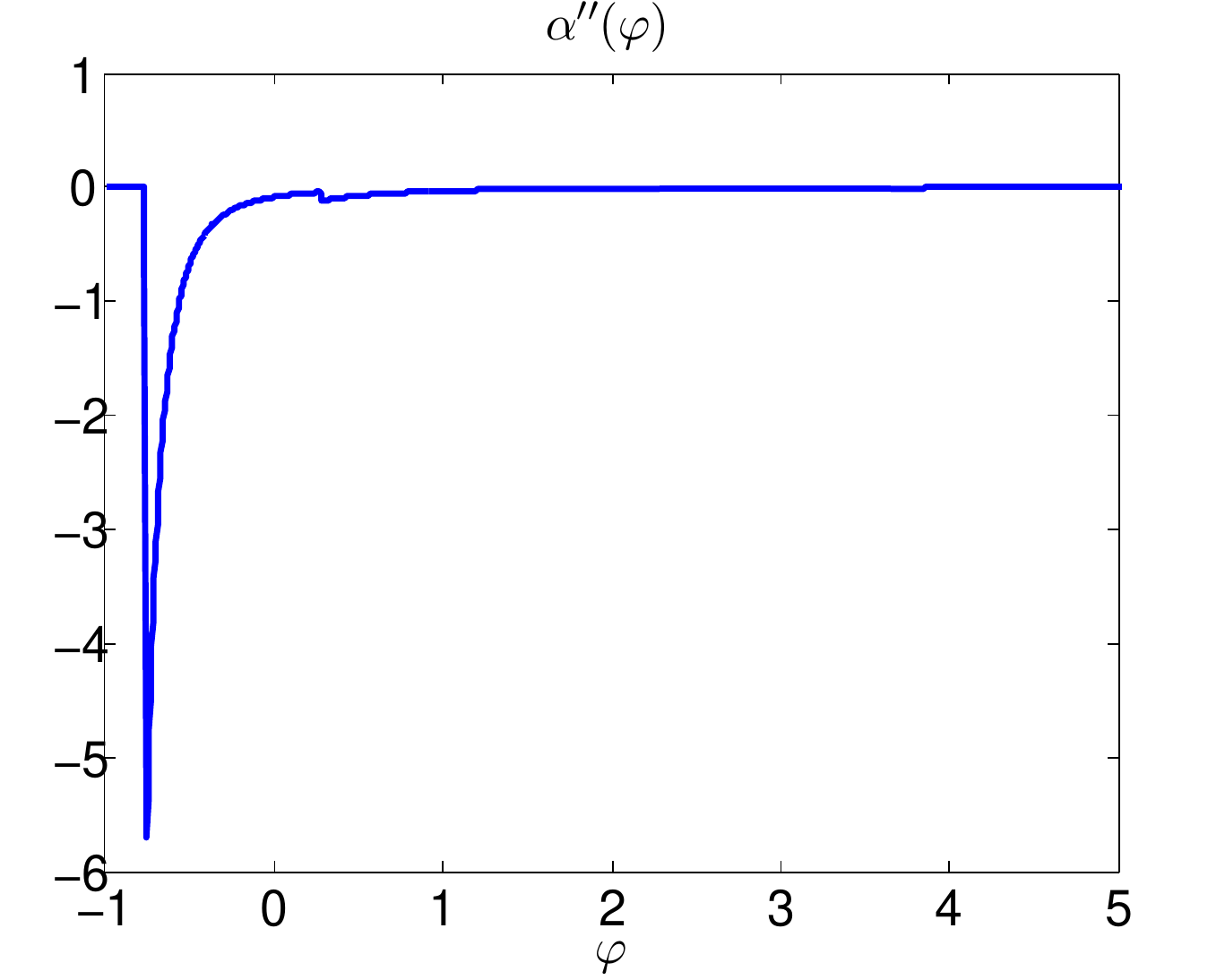}
    \caption{Illustration of function $\alpha$ and its first two derivatives with respect to $\varphi$ for Example \ref{exmp:portfolio1}.}
    \label{fig:alpha_alphader_alphaderder}
\end{figure}
\end{example}

In what follows, we will use the notation $\partial_x\alpha$ for the total differential of the function $\alpha(x,t,\varphi(x,t))$, i.e. 
$\partial_x\alpha (x,t,\varphi(x,t)) := \alpha^\prime_x(x,t,\varphi(x,t)) + \alpha^\prime_\varphi(x,t,\varphi(x,t))\, \partial_x\varphi(x,t)$,
where $\alpha^\prime_x$ and $\alpha^\prime_\varphi$ denote partial derivatives of the function $\alpha=\alpha(x,t,\varphi)$ with respect to $x$ and $\varphi$, respectively.

In \cite{KilianovaSevcovicANZIAM} we investigated a class of problems with the drift and variance functions discussed in the previous example. The goal of the next theorem is to extend the Riccati transformation methodology (see \cite[Theorem 3.3]{KilianovaSevcovicANZIAM}) to a broader class of stochastic processes of the form (\ref{process_x}) with a general  drift and variance functions including in particular important applications in the worst case dynamic portfolio optimization studied by Kilianov\'a and Trnovsk\'a in \cite{KilianovaTrnovska}. 

\begin{theorem}\label{th-equiv}
Assume that $U:\R\to\R$ is a differentiable increasing function. Then an increasing function $V(x,t)$ is a solution to the Hamilton-Jacobi-Bellman equation (\ref{eq_HJB})--(\ref{init_eq_HJB}) if and only if the transformed function $\varphi= -\partial_x^2 V(x,t)/\partial_x V(x,t)$ is a solution to the quasi-linear parabolic PDE:
\begin{eqnarray}
&&\partial_t \varphi + \partial_x\left(\partial_x\alpha(\cdot,\varphi) - \alpha(\cdot,\varphi)\varphi\right) = 0, \quad (x,t)\in\R\times[0,T), 
\label{eq_PDEphi}
\\
&&\varphi(x,T) = -U''(x)/U'(x),\quad x\in\R, \label{init_PDEphi}
\end{eqnarray}
and
\begin{equation}
V(x,t) = a(t) + b(t) \int_{x_0}^x e^{-\int_{x_0}^\xi \varphi(\eta,t) d\eta} d\xi
\label{Vexpression}
\end{equation}
where 
$b(t) = U^\prime(x_0) e^{-\int_t^T\omega(\tau) d\tau}$, $a(t) = U(x_0) - \int_t^T \gamma(\tau) b(\tau) d\tau$ and the functions $\gamma$ and $\omega$ are given by 
$\gamma(t) := \alpha(\varphi(x_0,t), x_0, t), \omega(t) := \partial_x\alpha(\varphi(x_0,t), x_0, t) - \alpha(\varphi(x_0,t), x_0, t) \varphi(x_0,t)$ where $x_0\in\R$ is a fixed real number. 
\end{theorem}

\noindent P r o o f. Let $V$ be a solution to the HJB equation (\ref{eq_HJB}) satisfying the terminal condition (\ref{init_eq_HJB}) and such that $\partial_x V(x,t)>0$ for each $(x,t)\in\R\times[0,T)$. Thus $V$ solves (\ref{eq_HJBtransf}), i.e. $\partial_t V = \alpha(x,t,\varphi)\, \partial_x V$ where $\varphi=-\partial^2_x V/\partial_x V$. 
Since
\[
\partial_t\varphi = -\frac{\partial^2_x\partial_t V}{\partial_x V} + \frac{\partial^2_x V \partial_x\partial_t V}{(\partial_x V)^2}
= -\frac{\partial^2_x\partial_t V}{\partial_x V} - \varphi\frac{\partial_x\partial_t V}{\partial_x V}, 
\]
\[
\partial^2_x V = -\varphi \partial_x V,\quad\hbox{and}\quad 
\partial^3_x V = - \partial_x(\varphi \partial_x V) = (\varphi^2 - \partial_x\varphi) \partial_x V,
\]
it follows from the equation $\partial_t V - \alpha \partial_x V =0$ that $\varphi$ satisfies:
\begin{eqnarray*}
\partial_t\varphi &=& -\frac{1}{\partial_x V} \left( 
\partial^2_x\alpha\, \partial_x V + 2 \partial_x\alpha \, \partial^2_x V +\alpha\,\partial^3_x V + \varphi\partial_x\alpha\,\partial_x V +\varphi\alpha\,\partial^2_x V
\right)
\\
&=& -\frac{1}{\partial_x V} \left( 
\partial^2_x\alpha\, \partial_x V -\varphi \partial_x\alpha\,\partial_x V +\alpha(\varphi^2-\partial_x\varphi)\partial_x V - \varphi^2\alpha\,\partial_x V
\right)
\\
&=& -\partial^2_x\alpha -\varphi \partial_x\alpha-\alpha\,\partial_x\varphi
= - \partial_x\left(\partial_x\alpha - \alpha\varphi\right).
\end{eqnarray*}
It means that the function $\varphi$ is a solution to the Cauchy problem (\ref{eq_PDEphi})--(\ref{init_PDEphi}). 

Moreover, by differentiating (\ref{eq_HJBtransf}) with respect to $x$ we obtain $\partial_t \partial_x V = \partial_x(\alpha \partial_x V) = \partial_x\alpha\, \partial_x V + \alpha \partial^2_x V
=  (\partial_x\alpha -  \alpha\,\varphi) \partial_x V$. Taking $x=x_0$ we conclude 
$\partial_t \partial_x V(x_0,t) = \omega(t) \partial_x V(x_0,t)$. As $\partial_x V(x_0,T) =U^\prime(x_0)$ we obtain $\partial_x V(x_0,t)= U^\prime(x_0) e^{-\int_t^T\omega(\tau)\d\tau}=b(t)$. Furthermore, as $\partial_t V(x_0,t) = \alpha(x_0,t,\varphi(x_0,t)) \partial_x V(x_0,t) = \gamma(t) b(t)$ and $V(x_0,T)= U(x_0)$ we obtain $V(x_0,t)=a(t)$. Since $\varphi(x,t)=-\partial^2_x V(x,t)/\partial_x V(x,t)$ we obtain
\begin{eqnarray*}
a(t) + b(t) \int_{x_0}^x e^{-\int_{x_0}^\xi \varphi(\eta,t) d\eta} d\xi
&=& a(t) + b(t) \int_{x_0}^x e^{\int_{x_0}^\xi \partial^2_\eta V(\eta,t)/\partial_\eta V(\eta,t) d\eta} d\xi
\\
&=& a(t) + \frac{b(t)}{\partial_x V(x_0,t)} \int_{x_0}^x \partial_\xi V(\xi,t)  d\xi =V(x,t), 
\end{eqnarray*}
as claimed. Now, if $\varphi$ solves (\ref{eq_PDEphi})--(\ref{init_PDEphi}) then the function $V(x,t)$ given by (\ref{Vexpression}) satisfies $-\partial^2_x V(x,t)/\partial_x V(x,t) = \varphi(x,t)$. Moreover, $V(x,T) = a(T) + b(T) \int_{x_0}^x e^{-\int_{x_0}^\xi \varphi(\eta,T) d\eta} d\xi = U(x_0) + U^\prime(x_0) \int_{x_0}^x e^{\int_{x_0}^\xi U''(\eta)/U'(\eta) d\eta} d\xi = U(x)$. The function $V(x,t)$ is increasing in the $x$ variable as $\partial_x V(x,t) = b(t) e^{-\int_{x_0}^x \varphi(\xi,t) d\xi} >0$. Now, as $da/dt = \gamma b$ and $db/dt = \omega b$ and $\partial_x V(x,t) = b(t) e^{-\int_{x_0}^x \varphi(\eta,t) d\eta}$  we obtain 
\begin{eqnarray*}
\partial_t V(x,t) &=& \frac{d a}{dt}(t) +  \int_{x_0}^x \frac{d b}{dt}(t) e^{-\int_{x_0}^\xi \varphi(\eta,t) d\eta} 
 - b(t)  e^{-\int_{x_0}^\xi \varphi(\eta,t) d\eta} 
\int_{x_0}^\xi \partial_t\varphi(\eta,t) d\eta
d\xi
\\
&=& \gamma(t)b(t)  + \omega(t) (V(x,t)-a(t)) 
 - \int_{x_0}^x \partial_\xi V(\xi,t) \left(\int_{x_0}^\xi \partial_t\varphi(\eta,t) d\eta\right)
d\xi.
\end{eqnarray*}
Since  $\partial_t\varphi = -\partial_x(\partial_x \alpha - \alpha\,\varphi)$ and 
\begin{eqnarray*}
\int_{x_0}^x \partial_\xi V  \left(\partial_\xi\alpha -\alpha\,\varphi \right)d\xi
&=& \alpha\partial_x V - \gamma(t) b(t)  + \int_{x_0}^x -\partial^2_\xi V \, \alpha - \partial_\xi V\, \alpha \varphi d\xi 
\\
&=&  \alpha\partial_x V - \gamma(t) b(t)
\end{eqnarray*}
we have $\int_{x_0}^x \partial_t\varphi(\eta,t) d\eta = -\left(\partial_x\alpha(x,t,\varphi(x,t)) -\alpha(x,t,\varphi(x,t))\,\varphi(x,t) \right) +  \omega(t)$. Hence $\partial_t V(x,t) \alpha(x,t,\varphi(x,t)) \partial_x V(x,t) = \alpha(x,t,-\partial^2_x V/\partial_x V) \partial_x V(x,t)$,
which means that $V(x,t)$ solves (\ref{eq_HJBtransf}) and, consequently, the HJB equation (\ref{eq_HJB})--(\ref{init_eq_HJB}).  \hfill$\diamondsuit$

\section{A parametric quadratic programming problem}
\label{sec:multiportf}

In this section we analyze qualitative properties of the optimal value function $\alpha(x,t,\varphi)$ defined by means of the convex optimization problem (\ref{eq_alpha_def}). By $C^{k,1}$ we denote the space of all functions whose $k$-th derivative is Lipschitz continuous. The following result is a generalization of \cite[Theorem 4.1]{KilianovaSevcovicANZIAM} for a more general drift and volatility functions. 

\begin{theorem}\label{smootheness}
Assume that the functions $(x,t,\bmtheta)\mapsto\mu(x,t,\bmtheta)$ and $(x,t,\bmtheta )\mapsto\sigma(x,t,\bmtheta)$ are $C^1$ smooth in $x, t$ and $\bmtheta$ variables, and such that the objective function 
$f(x,t,\varphi, \bmtheta):= - \mu(x,t,\bmtheta) + \frac{\varphi}{2} \sigma(x,t,\bmtheta)^2$
is strictly convex in the $\bmtheta$ variable for any  $\varphi\in(\varphi_{min}, \infty)$. 
Assume $\Delta\subset\mathcal{S}^n$ is a closed convex subset. Then the optimal value function $\alpha(x,t,\varphi)$ defined as in (\ref{eq_alpha_def}) is a $C^{1,1}$ continuous function for $x\in\R, t\in[0,T), \varphi>\varphi_{min}$. Moreover, $\varphi\mapsto\alpha(x,t,\varphi)$ is a  strictly increasing function, and 
\begin{equation}
\alpha^\prime_\varphi(x,t,\varphi) =\frac{1}{2} \sigma(x,t, \hat\bmtheta(x,t,\varphi))^2,  
\label{eq_alphader_vzorec}
\end{equation}
where $\hat\bmtheta(x,t,\varphi) \in \Delta\subset \mathcal{S}^n$ is the unique minimizer of (\ref{eq_alpha_def}) for $\varphi >\varphi_{min}$. Moreover, the function $\R\times[0,T)\times(\varphi_{min},\infty) \ni (x,t,\varphi) \mapsto \hat{\bmtheta}(x,t,\varphi) \in\R^n$  is Lipschitz continuous.
\end{theorem}

\noindent P r o o f. The mapping  $(x,t,\varphi) \mapsto \hat{\bmtheta}(x,t,\varphi) \in\R^n$ is continuous, which can be deduced directly from basic properties of strictly convex functions minimized over the compact convex set $\Delta\subset \mathcal{S}^n$.

The objective function $f(x,t,\varphi, \bmtheta):= - \mu(x,t,\bmtheta) + \frac{\varphi}{2} \sigma(x,t,\bmtheta)^2$ in (\ref{eq_alpha_def}) is assumed to be  strictly convex in  the variable $\bmtheta$. Thus there exists a unique minimizer $\hat\bmtheta\equiv \hat\bmtheta(x,t,\varphi)$  to (\ref{eq_alpha_def}). 
Moreover, $f^\prime_\varphi(x,t,\varphi,\hat{\bmtheta}(x,t,\varphi))=\frac{1}{2}\sigma(x,t,\varphi, \hat\bmtheta(x,t,\varphi)^2$ is continuous in $(x,t,\varphi)$ due to continuity of $\hat{\bmtheta}(x,t,\varphi)$. By the envelope theorem due to Milgrom and Segal \cite[Theorem 2]{milgrom_segal2002} the function $\alpha(x,t,\varphi)$ is differentiable for $(x,t,\varphi)\in\R\times[0,T)\times(\varphi_{min},\infty)$.

The function $f(x,t,\varphi,\bmtheta)$ is linear in $\varphi$ for any $(x,t,\bmtheta)$. Therefore it is absolutely continuous in $\varphi$ for any $\bmtheta$. Again, applying \cite[Theorem
2]{milgrom_segal2002}, we obtain $\alpha^\prime_\varphi(x,t,\varphi) =
f^\prime_\varphi(x,t,\varphi,\hat\bmtheta(x,t,\varphi)) = \frac12 \sigma(x,t,\varphi, \hat\bmtheta(x,t,\varphi))^2 >0$. Hence $\varphi\mapsto \alpha(x,t,\varphi)$ is a $C^1$ con\-ti\-nuous and increasing function for $\varphi>\varphi_{min}$. Local Lipschitz continuity of $\alpha^\prime_\varphi(x,t,\varphi)$ now follows from the general result proved by Klatte in \cite{Klatte} (see also Aubin \cite{Aubin}). According to \cite[Theorem 2]{Klatte} the function $\hat\bmtheta(x,t,\varphi)$ is Lipschitz continuous. Hence the derivative $\alpha^\prime_\varphi(x,t,\varphi) = \frac12 \sigma(x,t,\varphi, \hat\bmtheta(x,t,\varphi))^2$ is locally Lipschitz continuous, as well.
\hfill$\diamondsuit$

\begin{remark}
The function $\varphi\mapsto\alpha(x,t,\varphi)$ need not by $C^2$ smooth as it was shown by Kilianov\'a and \v{S}ev\v{c}ovi\v{c} in \cite{KilianovaSevcovicANZIAM}. If the set $\Delta$ is not convex then $\hat\bmtheta(x,t,\varphi)$ need not be even continuous and, consequently, $\alpha$ need not be $C^1$ smooth. 
\end{remark}

\section{Numerical approximation scheme}

\label{sec:numscheme}

In this section, we recall a semi-implicit numerical
method for solving the Cauchy problem proposed and analyzed in \cite{KilianovaSevcovicANZIAM}. The method  is based on a finite volume approximation scheme (cf. LeVeque
\cite{LeV}) combined with a nonlinear equation iterative solver method  proposed by Mikula and K\'utik in \cite{KutikMikula}. Equation (\ref{eq_PDEphi}) belongs to a wide class of quasi-linear
parabolic equations of the general form:
\begin{equation}
\partial_t \varphi + \partial_{x}^2 A(x,t,\varphi) + \partial_{x} B(x,t,\varphi)+ C(x,t,\varphi) = 0, \quad x\in\R, t\in[0,T),
\label{eq_numerics_general}
\end{equation}
satisfying a terminal condition at $t=T$. Specifically for (\ref{eq_PDEphi}), where $
A(x,t,\varphi)=\alpha(x,t,\varphi),\quad  B(x,t,\varphi)= - \alpha(x,t,\varphi)\varphi(x,t), \quad  C \equiv 0$. We transform the equation
from backward time to a forward one via $\widetilde\varphi(x,
\tau) := \varphi(x, T-t)$. We obtain
\begin{equation}
\partial_\tau \widetilde\varphi(x,\tau)
=
 \partial_{x}^2 \widetilde A( x,\tau,\widetilde\varphi) + \partial_{x} \widetilde B( x,\tau,\widetilde\varphi)+ \widetilde C( x,\tau,\widetilde\varphi),
\quad\hbox{for any}\ x\in\R, \tau\in(0,T], \label{eq_num}
\end{equation}
with an initial condition
$\widetilde\varphi(x,0)=\widetilde\varphi_0(x) \equiv
\varphi(x,T)$. We have 
\[
\widetilde A(x,\tau,\varphi)=\alpha(x,T-\tau,\varphi),\quad  \widetilde B(x,\tau,\varphi)= - \alpha(x,T-\tau,\varphi)\varphi(x,T-\tau), \quad  \widetilde C \equiv 0.
\]
For convenience, we shall drop the $\widetilde{\phantom{a}}$ sign in the following, but we shall keep in mind that we work with the transformed functions.

Let us consider a bounded computational domain $[x_L, x_R]$ and spatial discretization points $x_i = x_L+ i h$ for
$i=0,\cdots,n+1$ where $h=(x_R-x_L)/(n+1)$. We have $x_0 =x_L$ and
$x_{n+1}=x_R$. Inner mesh points $x_i$, $i=1,\cdots,n$, are the
centers of the finite volumes cells $(x_{i - \frac{1}{2}}, x_{i
+\frac{1}{2}})$, which we will denote as $(x_{i-}, x_{i+})$ for simplicity. We
have $h=x_{i+} - x_{i-}$. Let the discretized time steps be $\tau^j=j k, j=0, \cdots,
m$, where $k=T/m$ and $m$ is the number of time steps in the considered time domain. Integrating equation (\ref{eq_num})
over finite volumes, applying the midpoint rule on the left-hand
side integral and approximating the time derivative by forward
finite difference with the time step $k$, we arrive at a system of equations:
\begin{equation}
\varphi_i^{j+1}  = \frac{k}{h}(I_1 + I_2) + \varphi_i^j\,, \quad
i=1, \cdots ,n,\ j=0, \cdots, m,
\end{equation}
\begin{eqnarray}
I_1 &=&\int_{x_{i-}}^{x_{i+}} \partial_x (\partial_x A(
x,\tau,\varphi) + B(x,\tau,\varphi)) {\d}x
=  [A^\prime_x + A^\prime_{\varphi} \partial_x\varphi + B]_{x_{i-}}^{x_{i+}},
\nonumber
\\
I_2 &=& \int_{x_{i-}}^{x_{i+}} C(x,\tau,\varphi) {\d}x
= h C(x_i,\tau,\varphi_i)\,.
\label{eq_defI1}
\end{eqnarray}
Depending on whether the above integrals are being computed on the
$j$-th or the $(j+1)$-th time layer, we obtain different approximations.
Before we specify how we treat this, we will use the symbol $^\star$ to denote either of $j$ or $j+1$.  If we denote
\begin{eqnarray*}
&& D_{i\pm}^\star =
A^\prime_{\varphi}(x,\tau,\varphi) |_{{x_{i \pm}}, \tau^\star,\varphi_{i
\pm}^\star}, \quad E_{i\pm}^\star =
A^\prime_{x}(x,\tau,\varphi) |_{{x_{i \pm}},\tau^\star,\varphi_{i \pm}^\star},
\\
&& F_{i\pm}^\star = B(x,\tau,\varphi)|_{
{x_{i \pm}}, \tau^\star,\varphi_{i \pm}^\star}, \quad
\partial_x \varphi|_{i\pm}^\star = \partial_x \varphi(x,\tau)|_{{x_{i \pm}},
\tau^\star}\,.
\end{eqnarray*}
and approximate the derivatives of $\varphi$ by 
\[
\partial _x \varphi|_{i+}^\star \approx
\frac{\varphi(x_{i+1},\tau^\star)-\varphi(x_i,\tau^\star)}{h},
\quad
\partial _x \varphi|_{i-}^\star \approx
\frac{\varphi(x_i,\tau^\star)-\varphi(x_{i-1},\tau^\star)}{h}.
\]
Next we apply a semi-implicit numerical scheme to compute a solution at the new time layer $j+1$. We take the terms $D_{i\pm}^\star, E_{i\pm}^\star, F_{i\pm}^\star$ from the previous time layer with $\star=j$ and the term $\partial_x \varphi|_{i\pm}^\star$ from the new layer with $\star=j+1$. Rearranging the new layer terms to the left-hand side and the old-layer terms to the right-hand side, we arrive at
\begin{eqnarray*}
-\frac{k}{h^2}D_{i+}^j \varphi_{i+1}^{j+1}
&+& (1+\frac{k}{h^2}(D_{i+}^j+D_{i-}^j)) \varphi_i^{j+1} -
\frac{k}{h^2} D_{i-}^j \varphi_{i-1}^{j+1} \\
&=& \frac{k}{h}(I_2^j + E_{i+}^j- E_{i-}^j + F_{i+}^j - F_{i-}^j)
+ \varphi_i^j\,,
\end{eqnarray*}
which is a tridiagonal system which can be effectively solved by the Thomas algorithm.

As boundary conditions we use the Robin condition on the left boundary $x_L$ and the Neumann condition on the right boundary $x_R$. More precisely, 
$$
\partial_x \varphi(x,\tau) = 1+ \varphi(x,\tau) \text{ at } x=x_L, \qquad \partial_x\varphi(x,\tau) = 0 \text{ at } x=x_R,
$$
for all $\tau \in (0,T]$. The boundary conditions follow from the asymptotic behavior of equation (\ref{eq_PDEphi}) for $x\to\pm \infty$. After discretization, boundary conditions take the form
$$
\varphi_0^{j} = \varphi_1^j/(1+h) - h/(1+h), \qquad \varphi_{n+1}^j (\tau) = \varphi_n^j(\tau).
$$

\section{Dynamic portfolio optimization with constant and decreasing risk aversion}
\label{sec:portfolio_example}

In this section we apply the numerical scheme from the previous section to an example of dynamic stochastic portfolio optimization. We use the scheme to find optimal portfolio weights as a result of expected utility maximization. Choosing a specific utility function is directly connected to determining the risk aversion of the investor. It can be measured by the well known Arrow-Pratt coefficient of absolute (or relative) risk aversion \cite{Arrow, Pratt}. The terminal condition \eqref{init_PDEphi} is the absolute risk aversion coefficient corresponding to the utility function $U$. Therefore, changing  the risk aversion changes the terminal condition for the quasi-linear PDE for $\varphi$, and so changes the optimal solution. It can be considered natural to assume that investors have decreasing risk aversion with increasing wealth: the more wealthy the investor is, the more open they are for risky positions with a possibility of achieving a higher return. We shall investigate how the results (in terms of return and riskiness) change when moving from an investor with a constant risk aversion to one with a decreasing risk aversion. 

\subsection{Data and parameters}

We consider a multi-period stochastic portfolio optimization problem with $n=30$ assets contained in the German DAX 30 index. We shall solve problem \eqref{maxproblem} for two exponential utility functions, the first one having a constant absolute risk aversion (CARA) and the other one with decreasing absolute risk aversion (DARA): 
\begin{eqnarray}
\text{CARA: } U(x) &=& - e^{-ax}, a = const., \label{eq:utility_CARA} \\
\text{DARA: } W(x) &=& \begin{cases} - e^{-a_0 x} - c^*, & x \le x^\ast, \\ 
- (a_0/a_1) e^{-a_1 x + (a_1-a_0)x^\ast}, & x > x^\ast, \end{cases} 
\label{eq:utility_DARA}
\end{eqnarray}
where $c^*=e^{-a_0 x^*}(a_0-a_1)/a_1$ is a constant and $a_0>  a_1>0$ and $x^\ast \in \mathbb{R}$ is a point at which the risk aversion changes. The DARA $C^1$ continuous function $W$ represents an investor with a non-constant, decreasing risk aversion: the higher the wealth, the lower their risk aversion and hence the higher  exposition of the portfolio to more risky assets. With regard to the paper \cite{Post} by Post, Fang and Kopa, the piece-wise exponential DARA utility functions play the important role in the analysis of decreasing absolute risk aversion stochastic dominance introduced by Vickson \cite{Vickson}. We note that the coefficients of absolute risk aversion of the above utility functions are $-U^{\prime\prime}(x)/U^{\prime}(x) \equiv a$ and $-W^{\prime\prime}(x)/W^{\prime}(x) \equiv a_0$ ($x\le x^*$) or $a_1$ ($x> x^*$) depending on the value of $x$. Our goal is to compare portfolio performance corresponding to constant risk aversion and a decreasing one. 

We used the following parameter values: regular inflow into the portfolio $\varepsilon = 1$ (per unit of time which is set to be one year), interest rate $r=0$, time horizon $T=10$ years. Numerical discretization parameters for solving the quasi-linear PDE \eqref{eq_PDEphi} for $\varphi$ are $h=0.05$, $k=0.05h^2$. We solved (\ref{eq_PDEphi}) in the bounded spatial domain $[x_L, x_R] = [\ln(0.01), 10]$. We computed the function $\alpha$ from \eqref{eq_alpha_def_quadratic} prior solving \eqref{eq_PDEphi} on the domain of $\varphi \in [-1, 15]$ with the discretization step $0.005$ in the $\varphi$ variable. Parameters $\bmmu$ and $\bmSigma$ for the set of 30 assets contained in the German DAX 30 index were calculated from historical data in the period of August 2010 - August 2012 (same data set as was used in \cite{KilianovaSevcovicANZIAM}) and an excerpt of them is summarized in Table \ref{tab:DAX5data}. Selected for Table are assets which in \cite{KilianovaSevcovicANZIAM} resulted in highest portfolio weights. We considered $a \in \{1,\cdots,15\}$ for the function $U$ and $a_0 \in \{4,\cdots,15\}$, $a_1 = a_0-3$, $x^\ast = 2$ for the function $W$. We chose a drop by 3 just  for illustration purposes, so the differences in results are more remarkable. In practice, one can choose an arbitrary choice of a position and size of a jump in the parameter $a$.

\begin{table}
\caption{
\small
Excerpt from the covariance matrix $\bmSigma^{part}$ and mean returns for six
stocks of the DAX 30 Index: Merck, Volkswagen, SAP, Fresenius
Medical, Linde, Fresenius. Based on historical data, August
2010--April 2012. Source: finance.yahoo.com, \cite{KilianovaSevcovicANZIAM}.
} 
\begin{center}
\scriptsize
\begin{tabular}{l || l | l | l | l | l |l || l }
$\bmSigma^{part}$ & Merck & VW & SAP & Fres Med & Linde & Fres &
Mean return
\\ \hline \hline
Merck & 1.6266 & -0.0155 & -0.0104 & -0.0146 & -0.0017 & -0.0033&
0.7315
\\ VW & -0.0155 & 0.1584 & 0.0345 & 0.0292 &
0.0569 & 0.0238 & 0.3413
\\ SAP & -0.0104 & 0.0345 & 0.0516
 & 0.0183 & 0.0240 & 0.0143&  0.1877
 \\ Fres Med & -0.0146 & 0.0292 &
0.0183
 & 0.0434 & 0.0227 & 0.0248 & 0.2202
\\ Linde &
-0.0017 & 0.0569 & 0.0240 & 0.0227 & 0.0530 & 0.0201 & 0.1932
\\ Fres &
-0.0033 & 0.0238 & 0.01430 & 0.0248 & 0.0201 & 0.0386 & 0.1351 \\ \hline
\end{tabular}
\end{center}
\label{tab:DAX5data}
\end{table}

Below we define a new portfolio performance measure. In order to illustrate its values and behaviour, we performed simulations of portfolio value evolution within the investment period $T$, starting from portfolio value $x_0 = 0$, with rebalancing based on optimal $\theta$ taking place in regular time instances $1/dt = 20$ with $dt = 0.05$.

\subsection{Risk-adjusted portfolio performance}

In order to compare the performance of the optimal portfolios with respect to a constant and decreasing risk aversion, we perform a set of 5000 simulations based on the optimal $\bmtheta$ for each of the utility functions and we evaluate their riskiness. To do so, let us first recall that if $v$ is a vector of realizations of a random variable obtained from Monte Carlo simulations, then the empirical estimates of the value-at-risk and conditional value-at-risk with level $\beta \in (0,1)$ (typically $\beta$ is between 0.01 and 0.05) are calculated as
$$
VaR_{\beta}(v) = F^{-1}(\beta), \qquad 
CVaR_{\beta}(v) = \mathbb{E}(v|v\le VaR_{\beta}(v)),  
$$
where $F^{-1}(\beta)$ is the quantile of the corresponding empirical distribution of the random variable $v$  (see li\-te\-ra\-tu\-re on risk measures, e.g. McNeil et al. \cite{McNeil} or Pflug and R\"omisch \cite{PflugKniha} and many others). In addition, we can define the so-called conditional value-at-risk deviation, which is the distance of $CVaR$ from the expected value, $CVaRD_{\beta}(v)  = \mathbb{E}(v) - CVaR_{\beta}(v)$,
(c.~f. \cite{PflugKniha}). This measure evaluates the amount of riskiness of investment in terms of down-side deviations of the return from its expected value. Some other possibilities for risk-adjusted measures can be found in Wiesinger \cite{Wiesinger}. 
To look at risk-adjusted performance, we can consider the standard Sharpe ratio \cite{Sharpe}
$$SR(v)= \frac{\mathbb{E}(v)-r}{StD(v)}
$$
or the Conditional Sharpe ratio which has first been used for performance measurement by Agarwal and Naik \cite{Argawal} and is defined as $SR_{CVaR_\beta}(v) = (\mathbb{E}(v)-r)/CVaR_{\beta}(v)$, see also Lin and Ohnishi \cite{Lin}. Another important measure for risk-adjusted performance of a portfolio is the Rachev ratio analyzed by Biglova {\it et al.} \cite{Rachev} (see also \cite{Farinelli} for survey of other portfolio performance measures).

The idea behind the standard Sharpe ratio is adjusting portfolio yield to the risk, which in this case is expressed by standard deviation of the considered return, characterizing how far random return realizations are spread around the mean value. As higher mean returns can often be naturally accompanied by higher values of $CVaR_\beta$, the Conditional Sharpe ratio could provide mystifying results in this sense. On the other hand,  $CVaRD_\beta$ measures certain distance from the mean value, so it is a more suitable replacement for standard deviation in the denominator of Sharpe ratio. In the light of this, we define a new measure of risk-adjusted performance. 

\begin{definition} Let $\beta>0$ be a given conditional value-at-risk level, $r$ the interest rate, and let $v$ be a random variable. We define the conditional value-at-risk deviation-based Sharpe ratio as
\begin{equation}
SR_{CVaRD_\beta}(v) = \frac{\mathbb{E}(v)-r}{\mathbb{E}(v) - CVaR_\beta(v)}.
\end{equation}
\end{definition}

We shall evaluate both the standard Sharpe ratio as well as its $CVaRD$-based version.

\subsection{Results}

\begin{figure}
    \centering
    \includegraphics[width=0.36\textwidth]{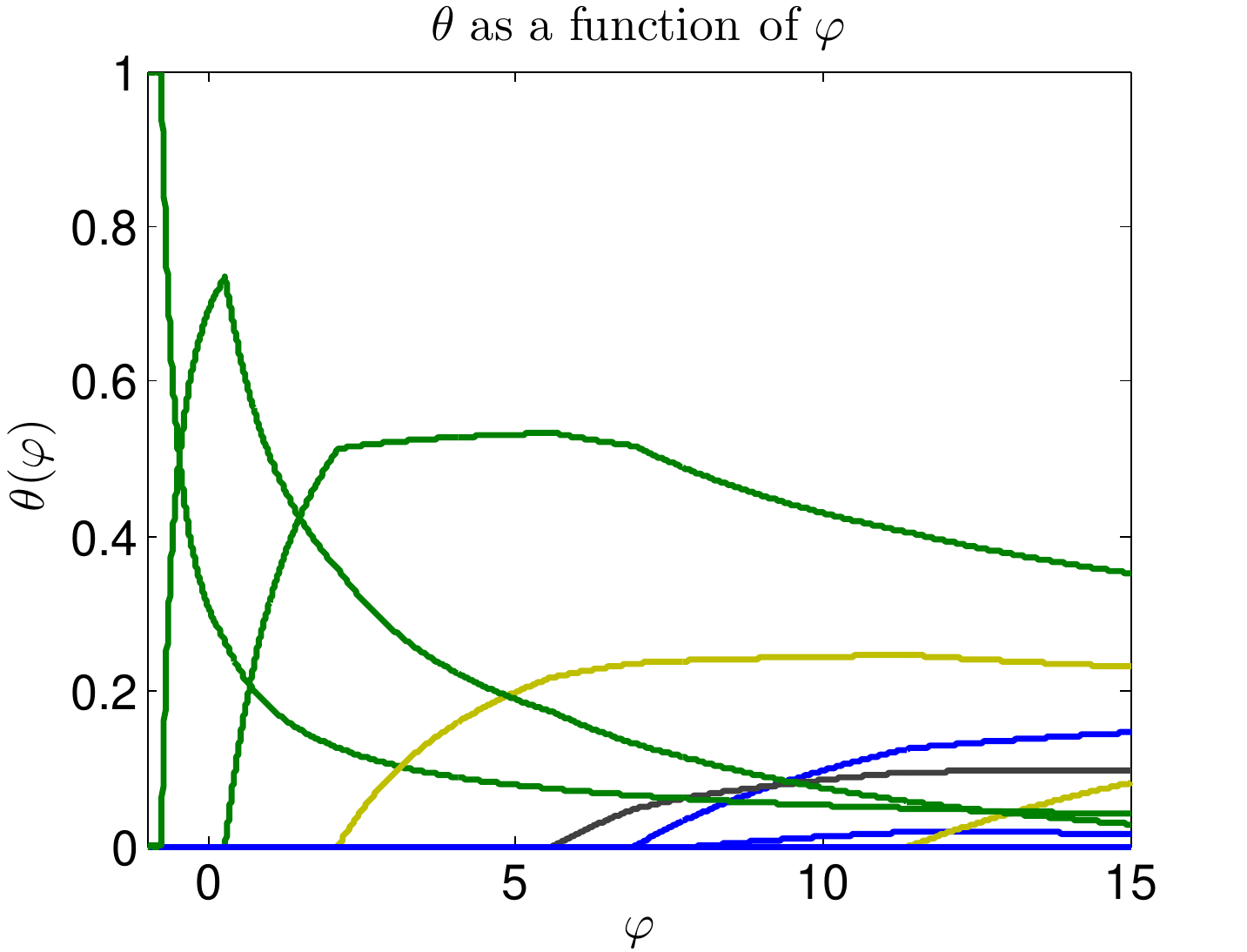} \\
    
    \includegraphics[width=0.36\textwidth]{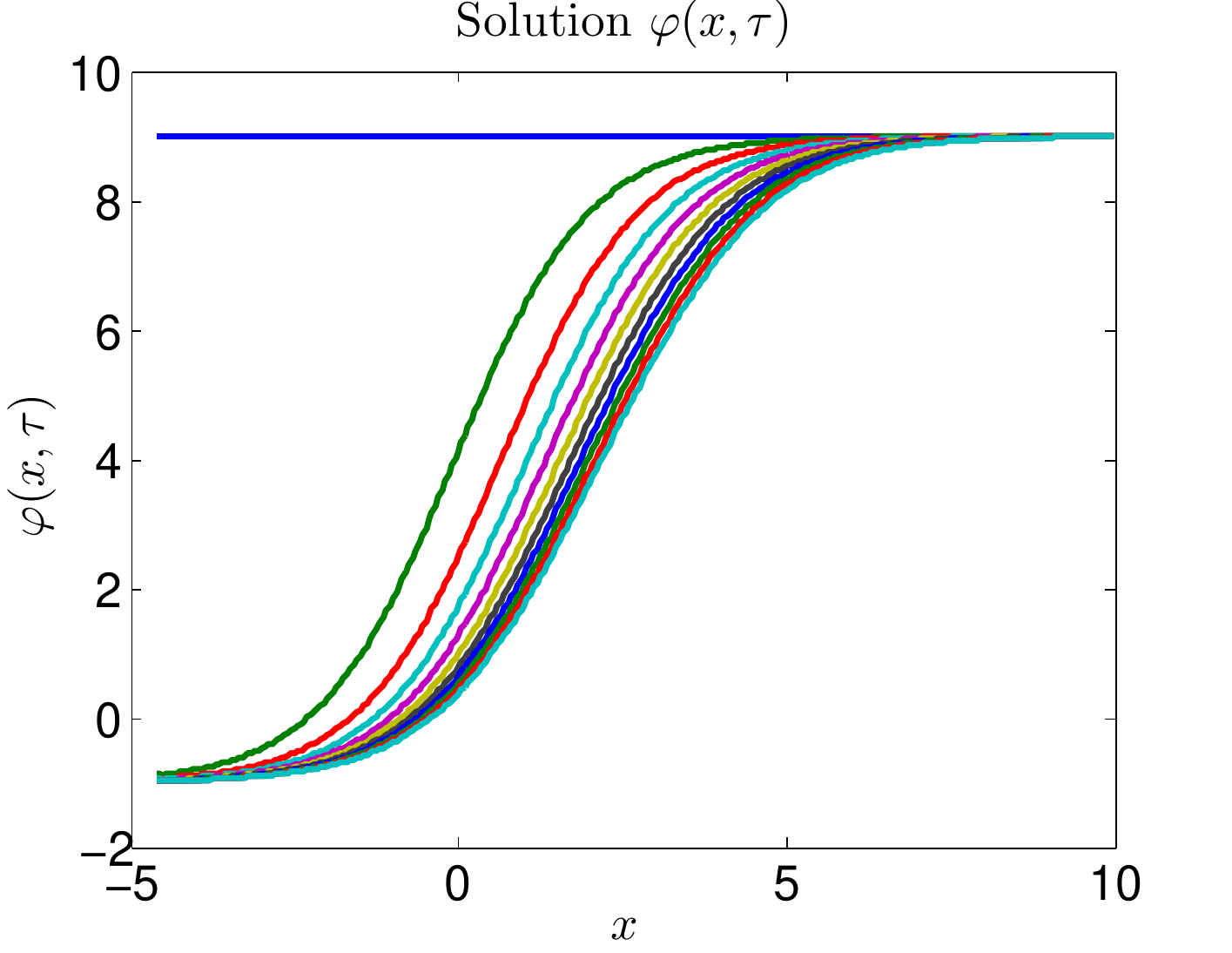} 
    \includegraphics[width=0.36\textwidth]{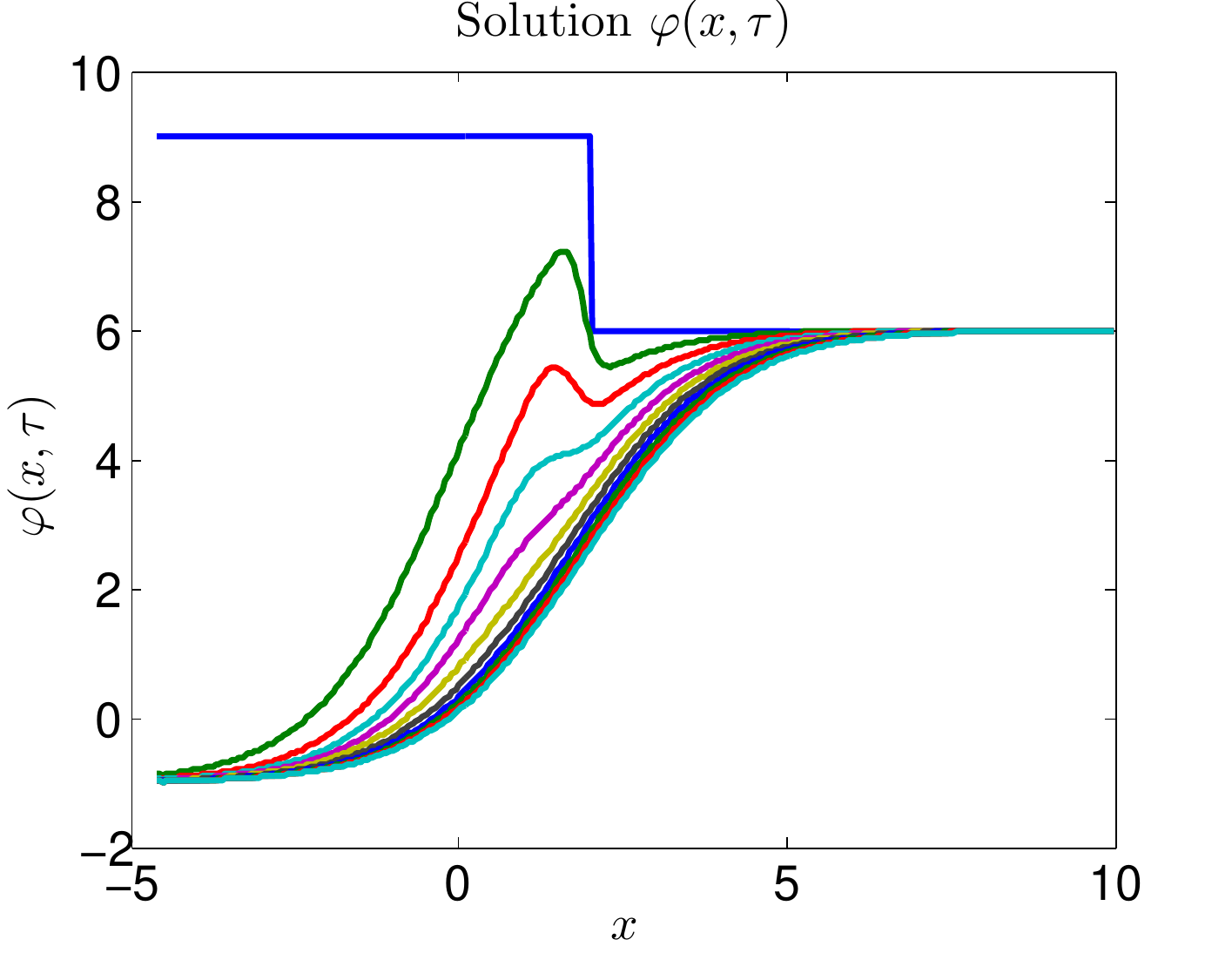}     

       \hglue-0.15truecm
       \includegraphics[width=0.36\textwidth]{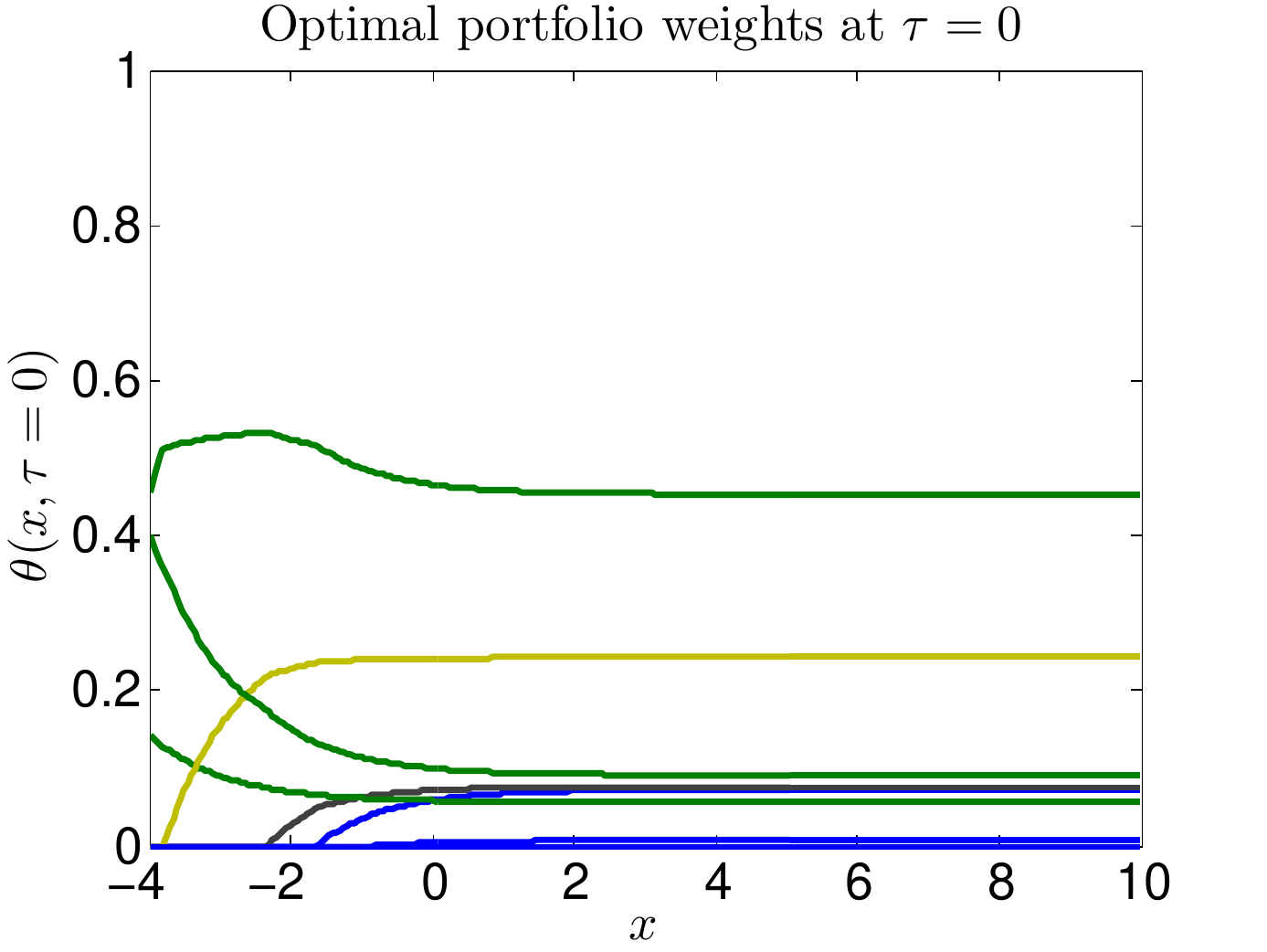} 
       \hglue-0.5truecm
       \includegraphics[width=0.36\textwidth]{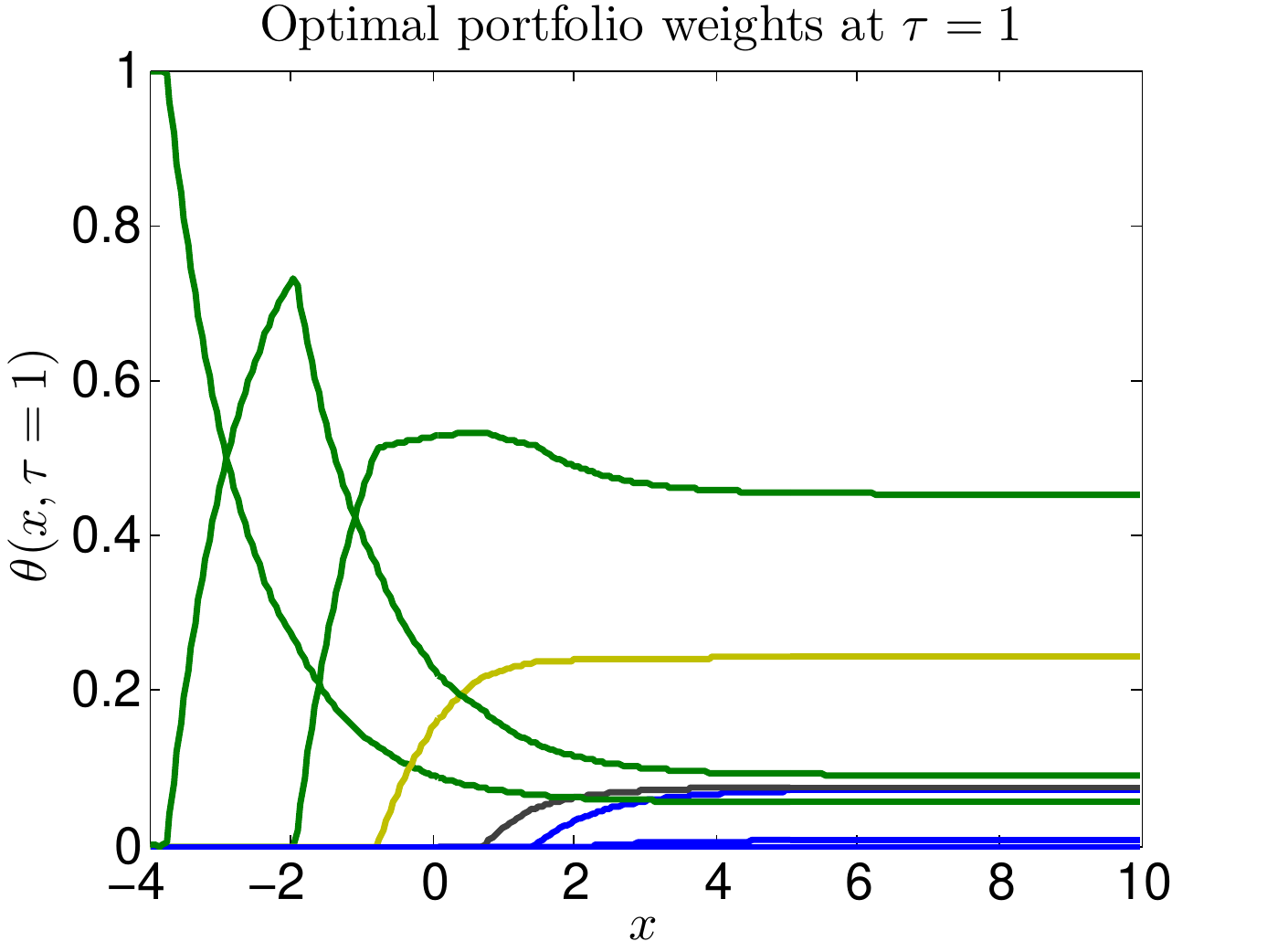} 
       \hglue-0.5truecm
       \includegraphics[width=0.35\textwidth]{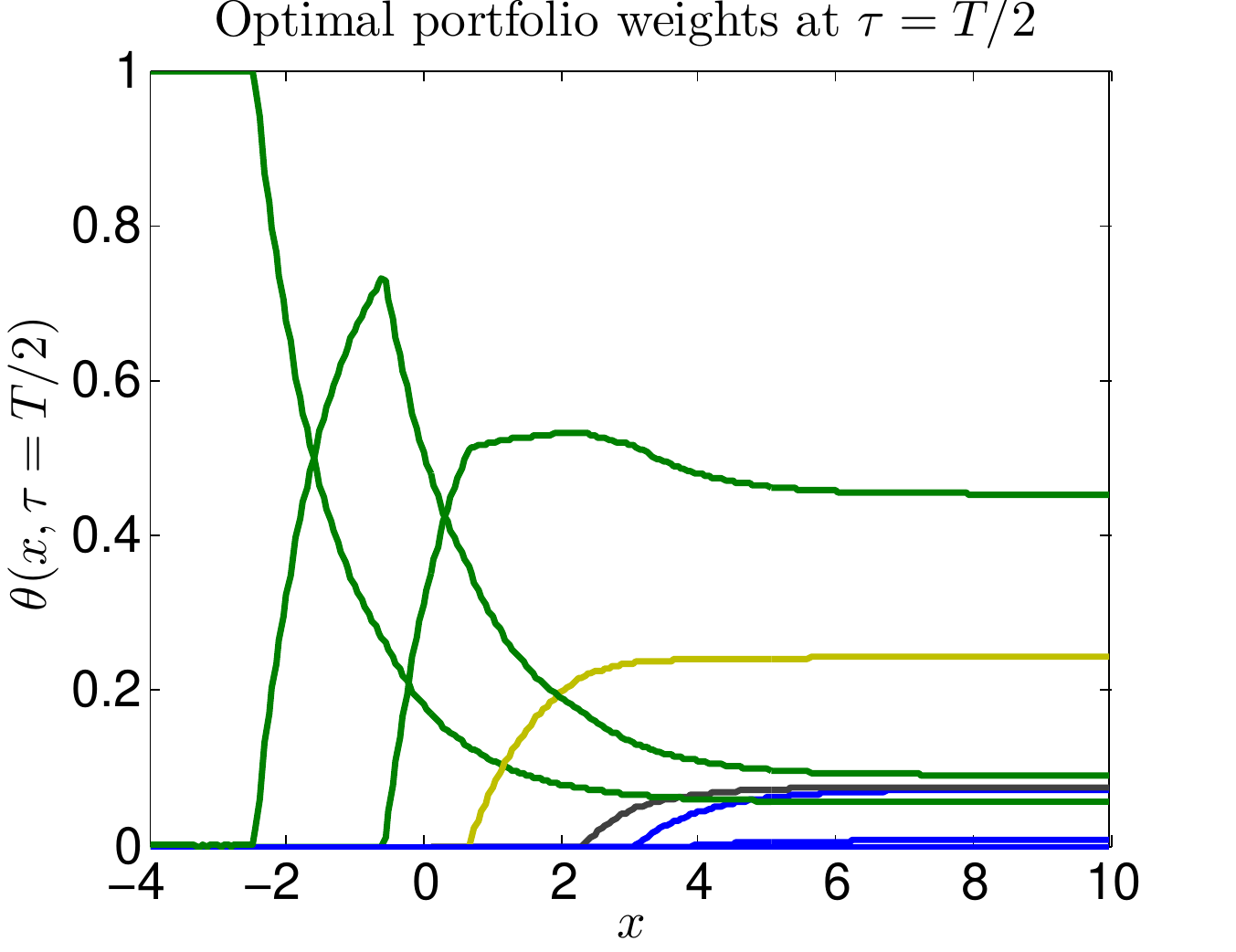}   \\ 
    
   \hglue-0.15truecm    \includegraphics[width=0.36\textwidth]{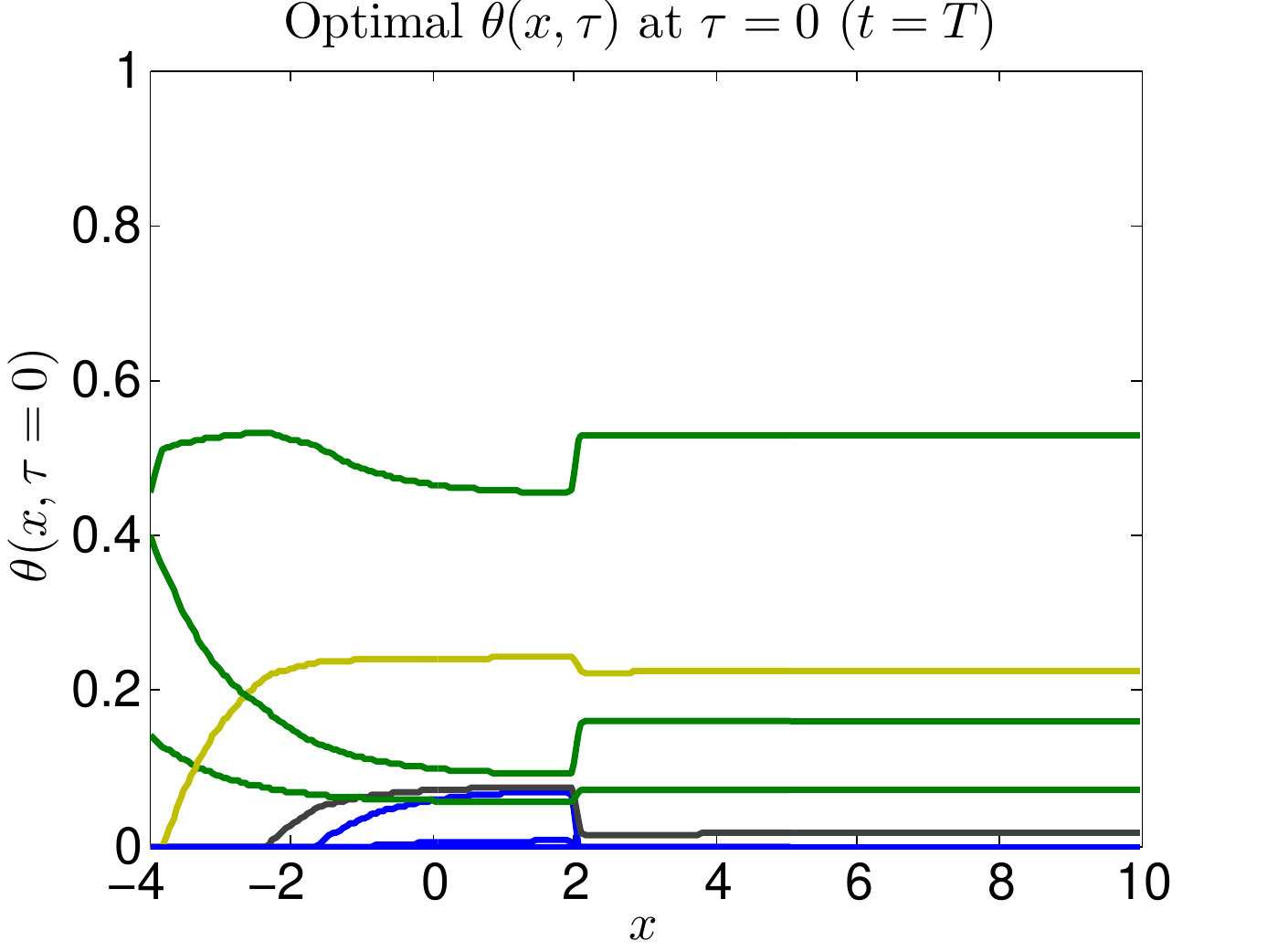}    \hglue-0.5truecm
    \includegraphics[width=0.36\textwidth]{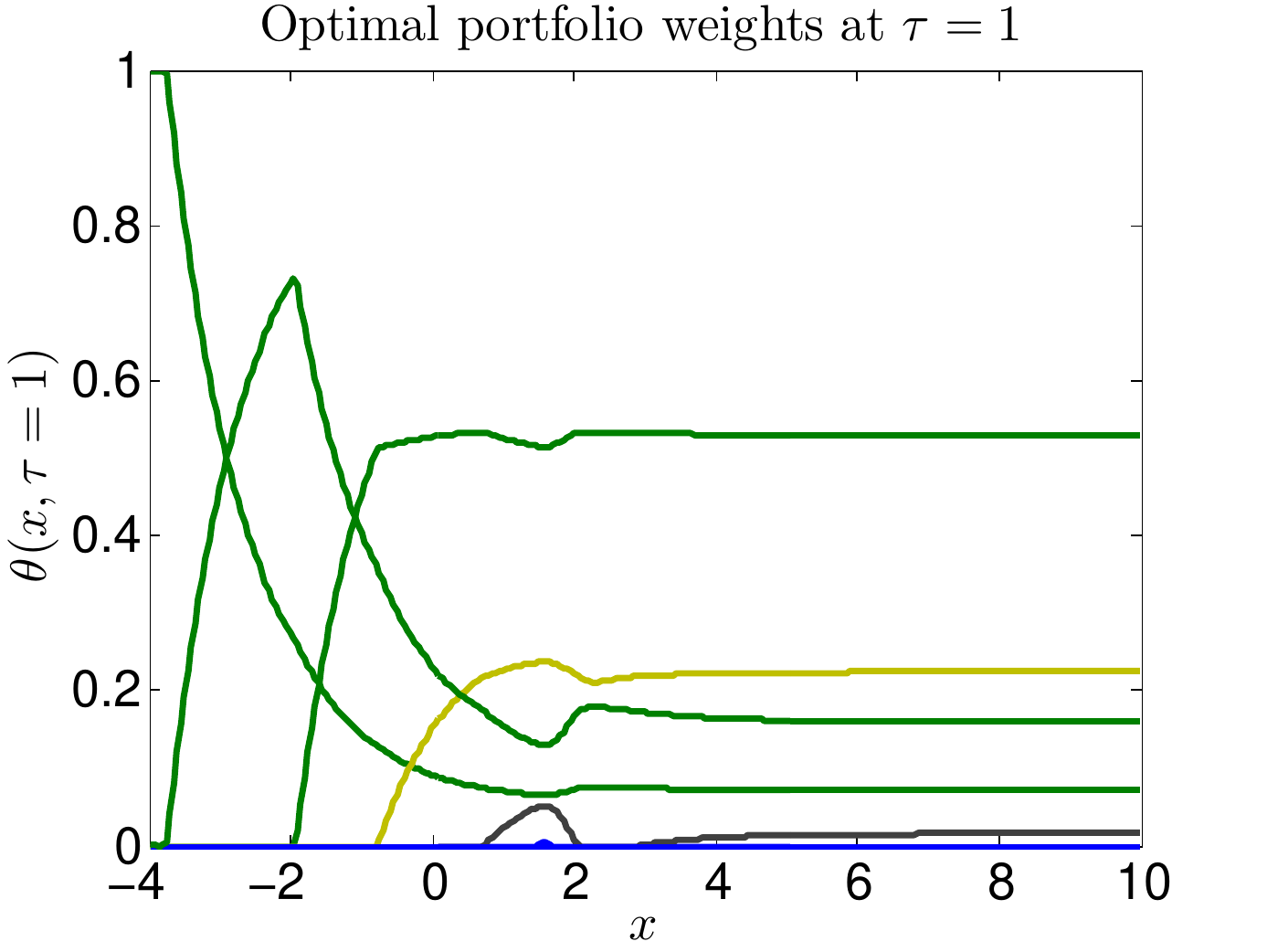}  
       \hglue-0.5truecm
       \includegraphics[width=0.35\textwidth]{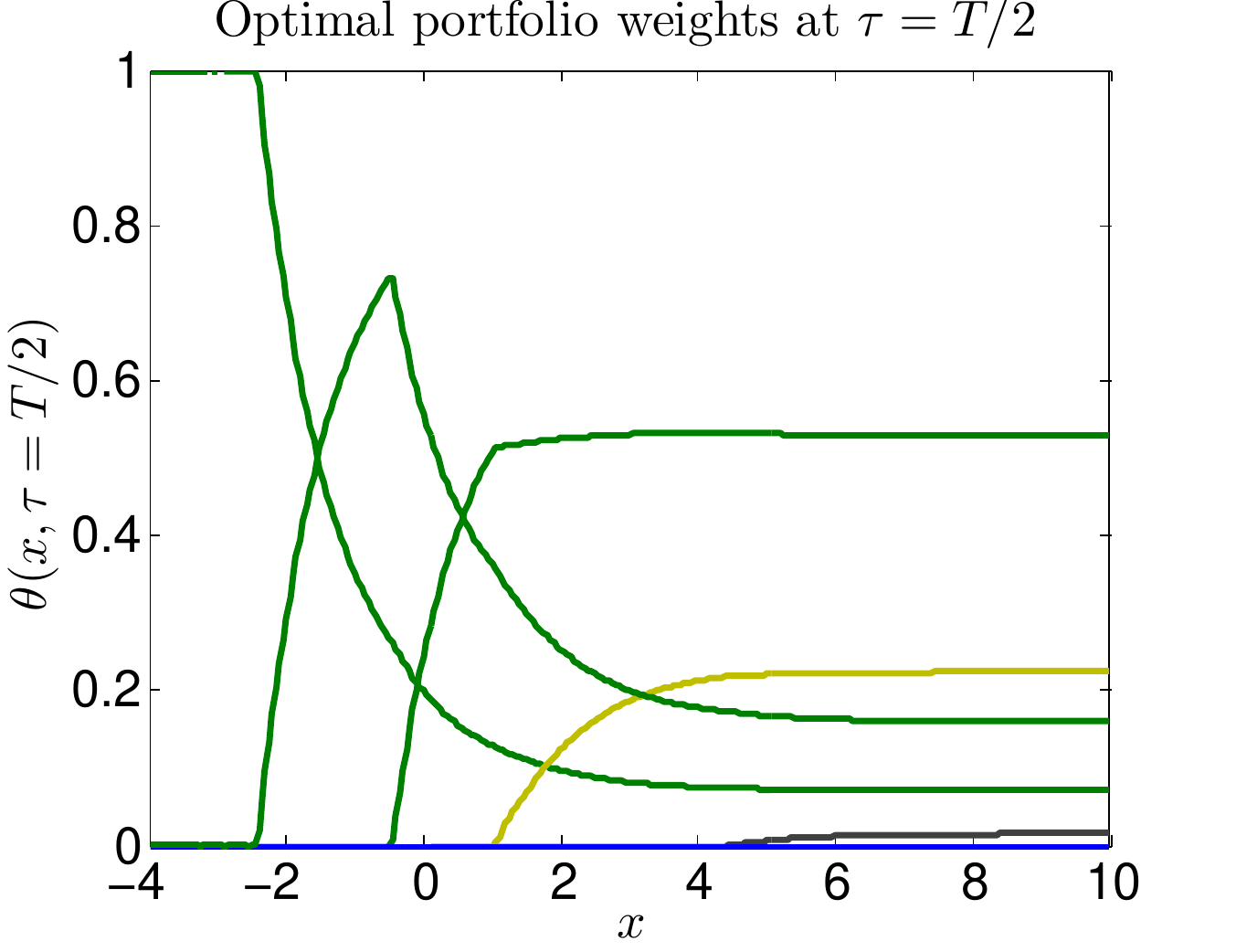} \\
    
    \caption{Top row: optimal $\bmtheta$ as a function of $\varphi$. Row 2: solutions $\varphi(x,\tau)$ for the utility function $U$ with constant $a=9$ (left) and for the DARA utility function $W$ with $a_0=9$, $a_1=6$, $x^\ast=2$ (right). Rows 3 and 4: optimal portfolio weights $\theta(x,\tau)$ at $\tau = 0, 1, T/2$, corresponding to utilities $U$ and $W$, respectively.}
    \label{fig:vysledky1}
\end{figure}

Figure \ref{fig:vysledky1} illustrates optimal weights $\theta(\varphi)$ obtained from solving \eqref{eq_alpha_def_quadratic}, which in our example is independent of the utility function, time $t$ or even wealth $x$. In the next row, the Figure displays solutions $\varphi(x,\tau)$ to the PDE \eqref{eq_PDEphi} for utility functions $U$ and $W$ at each integer time instance $\tau = \{0,1,\cdots, T\}$. The boundedness of the solution $\varphi(x,\tau)$  follows from the parabolic maximum principle (c.~f. \cite{KilianovaSevcovicANZIAM}). For a constant $a$ the function $\varphi(x,\tau)$ is increasing. However, for the step risk aversion, $\varphi(x,\tau)$ is not monotonous anymore. The bounds of $\varphi$ in connection to the graph of $\hat\bmtheta(\varphi)$ determine, which assets will enter the optimal portfolio with non-zero weights (see \cite{KilianovaSevcovicANZIAM} for more on this property). Note that, in our setting of the function $\alpha$, the optimal $\hat\bmtheta$ is independent of $x$ and $t$ variables. Rows 3 and 4 of the Figure show the optimal portfolio weights $\hat\bmtheta$ as functions of $x$ at selected time instances $\tau=0, 1$ and $T/2$. We can see that only few stocks enter the portfolio with non-zero weights.

Table \ref{tab:riskmeasures_const_a} summarizes the values of the mean $\mathbb{E}(x_T)$, standard deviation $StD(x_T)$ and  $CVaR_{\beta}(x_T)$ with $\beta=0.05$ for both utility functions $U$ and $W$, calculated from 5000 simulations of the process \eqref{process_x} using the optimal $\theta$ and the Euler-Maruyama numerical integration method. The values of these measures along with some others are presented also in Figure \ref{tab:riskmeasures_const_a} (left and middle). The results show that the utility function $W$ with a step-decrement in the risk aversion parameter $a$ yields higher mean returns but also higher values of risk measured by standard deviation or conditional value-at-risk. This makes it questionable whether this means a better or worse portfolio performance. In order to be able to answer this question, we computed the Sharpe ratios. 

\begin{figure}
    \centering
    \includegraphics[width=0.34\textwidth]{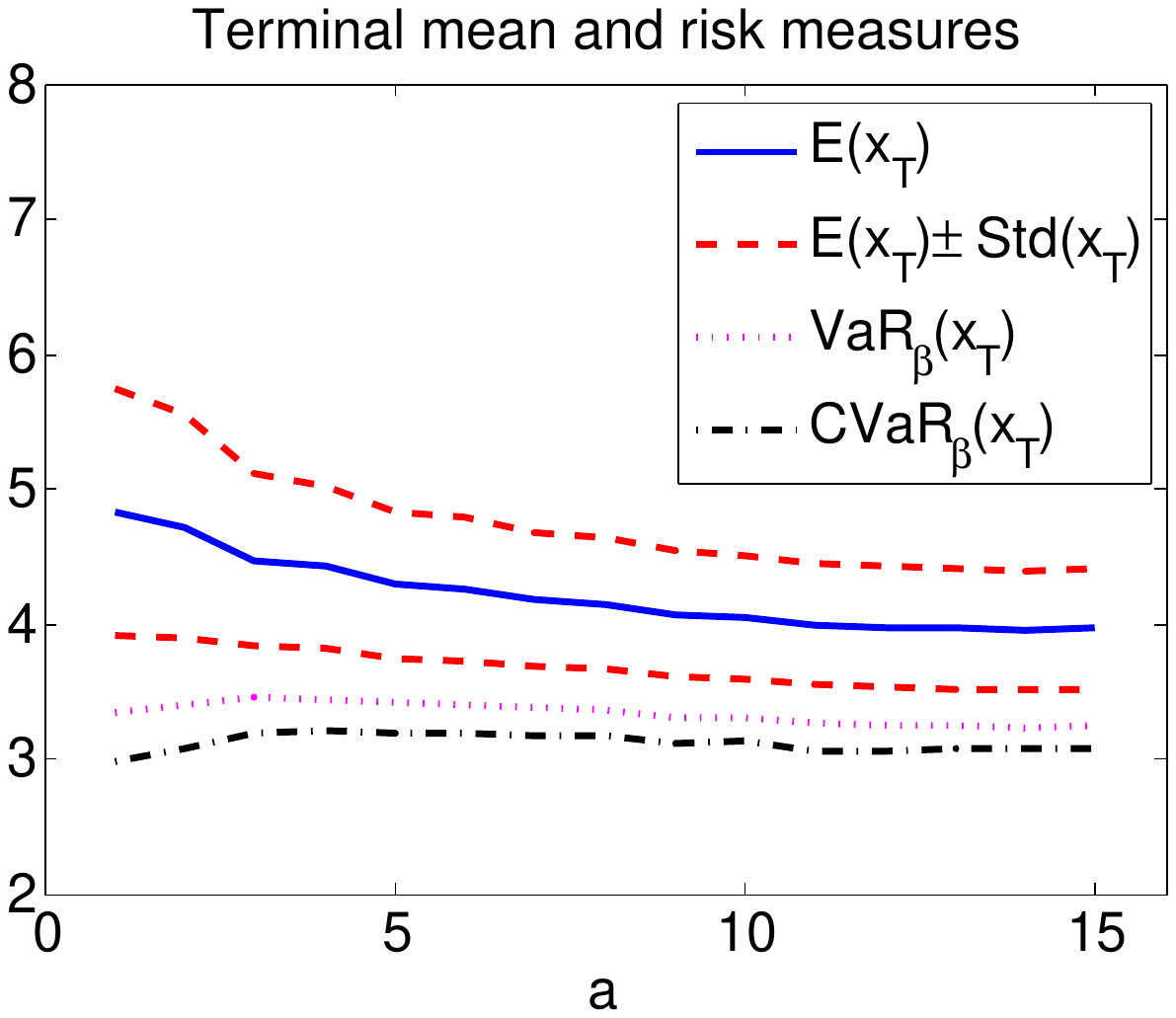}
       \hglue-0.5truecm
\includegraphics[width=0.35\textwidth]{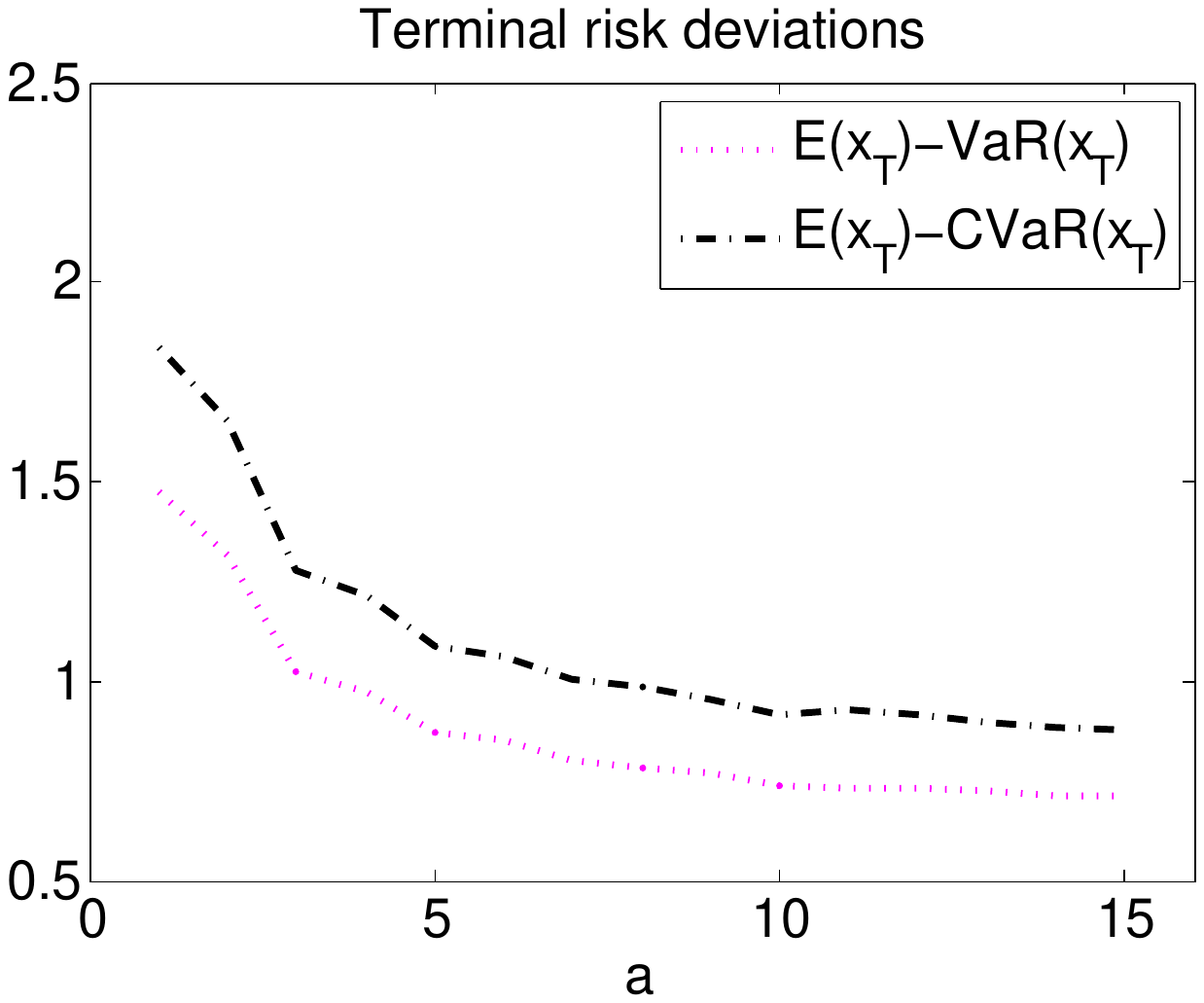}
   \hglue-0.5truecm
    \includegraphics[width=0.36\textwidth]{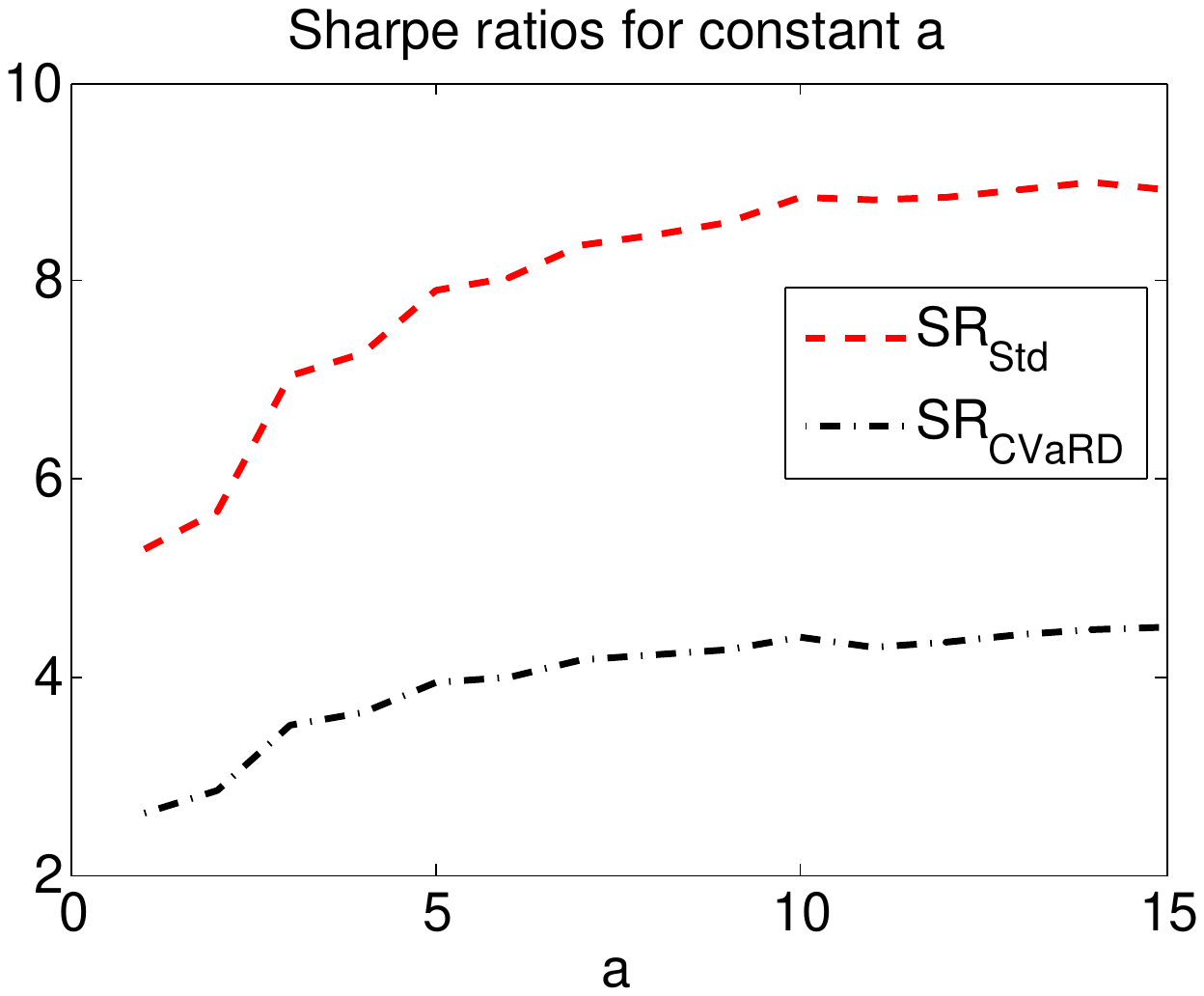}
    \includegraphics[width=0.34\textwidth]{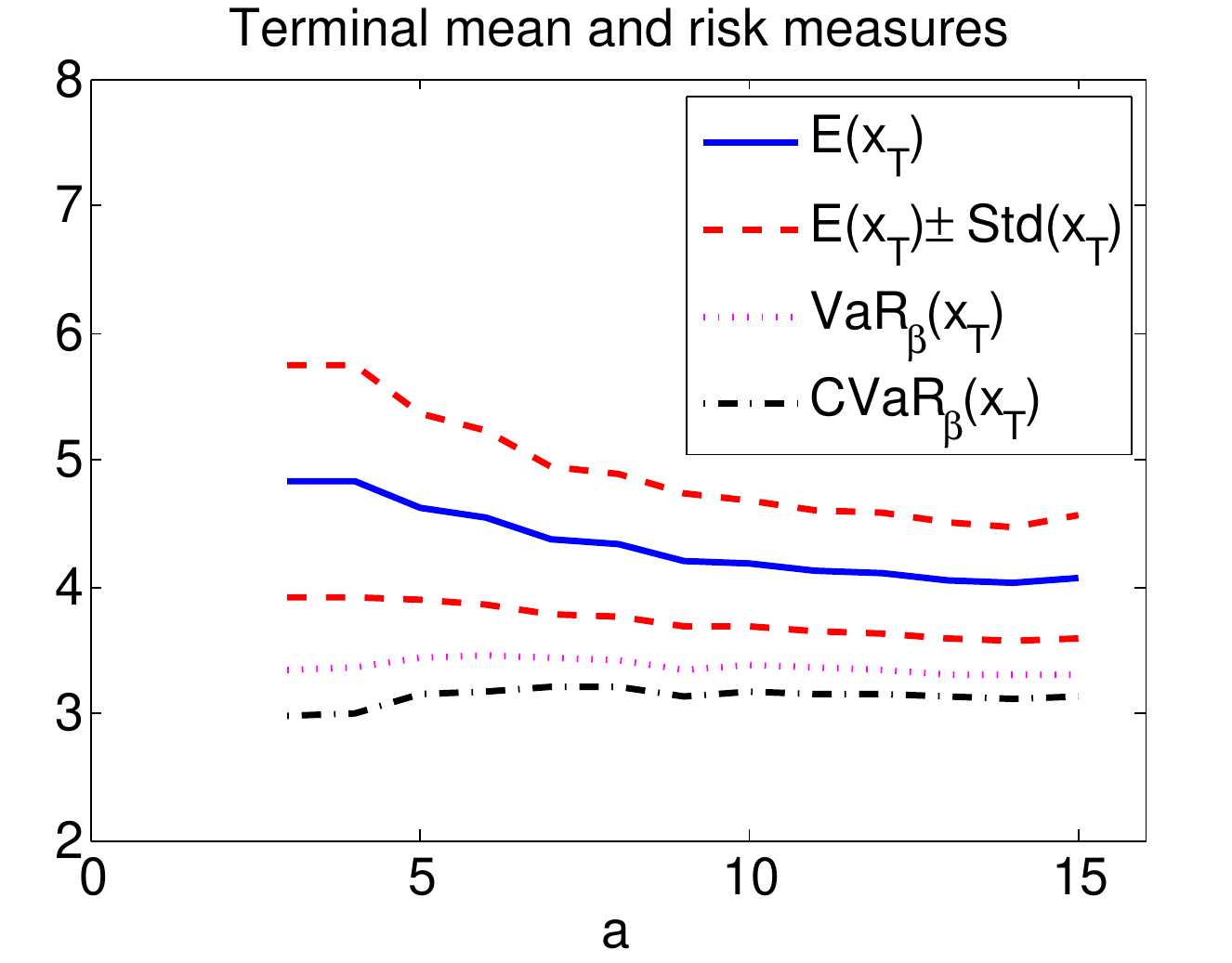}
   \hglue-0.5truecm
    \includegraphics[width=0.35\textwidth]{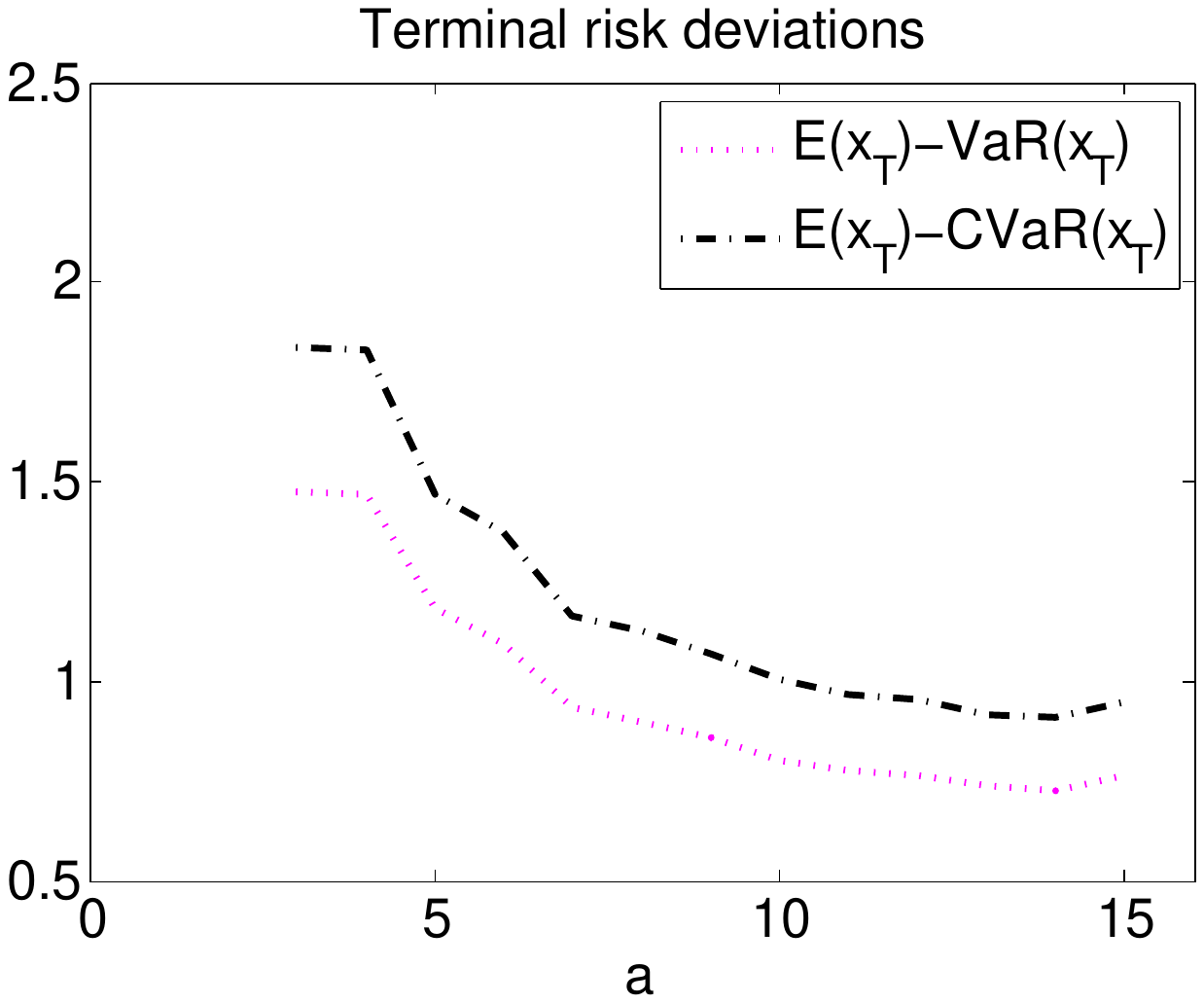}
   \hglue-0.5truecm
    \includegraphics[width=0.36\textwidth]{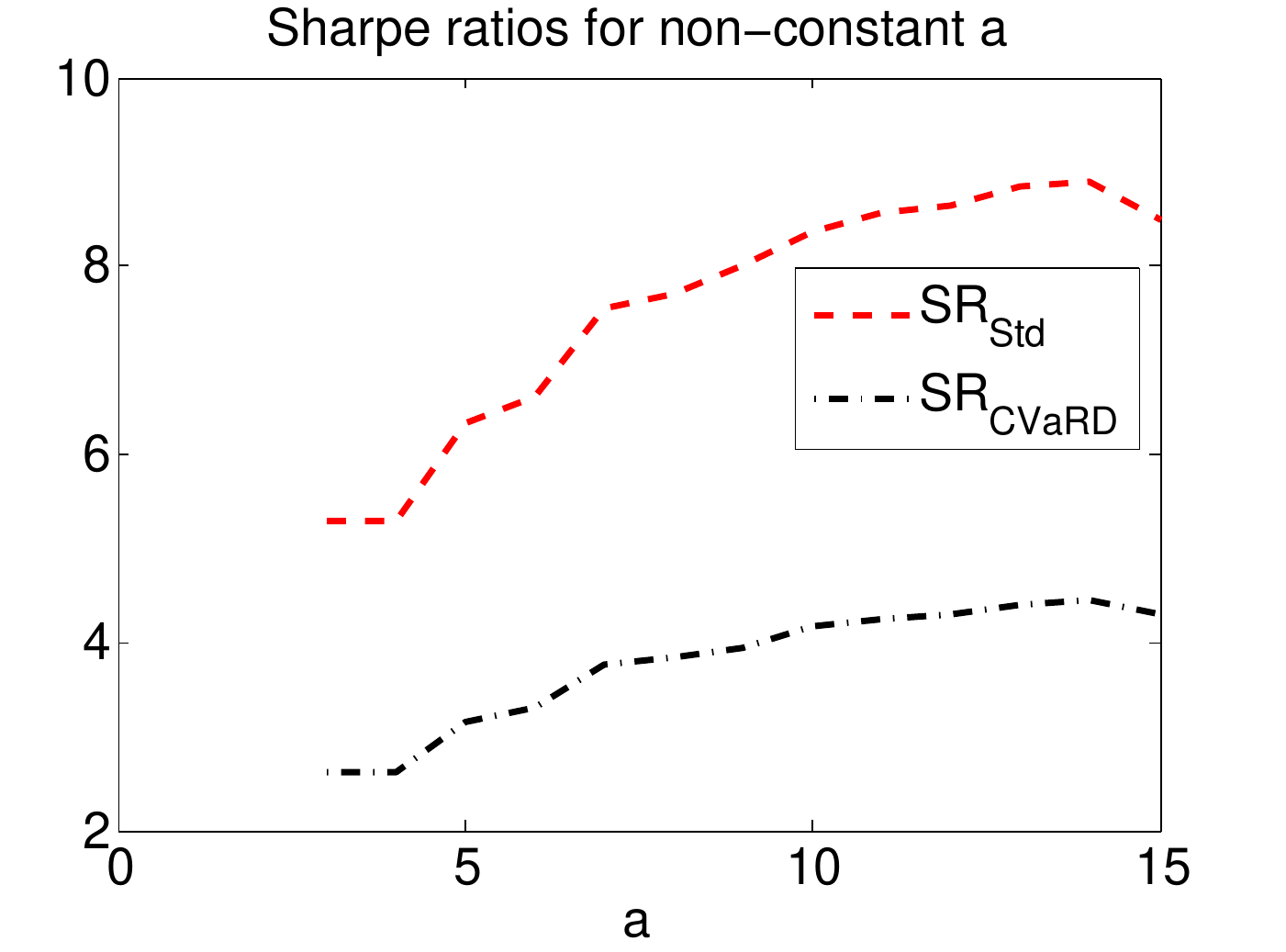}    
    \caption{Expected terminal wealth and associated risk (left) for various constant values of risk aversion parameter $a$ for utility function $U$ (or $a_0=a$, $a_1=a_0-3$ for $W$ in the second row), terminal risk deviations (middle) and corresponding $StD$-based and $CVaRD_\beta$-based Sharpe ratios (right). Results were obtained from 5000 simulations.}
    \label{fig:riskmeasures_const_a}
\end{figure}

Sharpe ratios for both the utility functions $U$ and $W$ are depicted in Figure \ref{fig:riskmeasures_const_a} (right) and summarized in Table \ref{tab:sharperatios}. We can see that both Sharpe ratios are lower for utility function $W$ with decreasing risk aversion, which illustrates (in this example, at least) that risk-adjusted performance of the associated portfolios is worse for lower risk aversion.

\begin{table}
\caption{Expected terminal wealth $x_T^U$ and $x_T^W$ for CARA/DARA utility functions $U$/$W$ and associated risk obtained from 5000 simulations for various constant values of risk aversion parameter $1\le a \le 15$ in $U$ and $4\le a_0\le 15$ and $a_1=a_0-3$ in $W$, $\beta=0.05$. 
\label{tab:riskmeasures_const_a}}
\scriptsize
\begin{center}
\begin{tabular}{l|ll|ll|ll}
$a$ & $\mathbb{E}(x_T^U)$ & $\mathbb{E}(x_T^W)$ & $StD(x_T^U)$ & $StD(x_T^W)$ & $CVaR_{\beta}(x_T^U)$  & $CVaR_{\beta}(x_T^W)$ \\ \hline \hline
1&	4.8268&	-- &0.91408& --	&2.9881     & -- \\
2&	4.7217&	-- &0.83286& --	&3.0682     & -- \\
3&	4.4761&	-- &0.63456& --    &3.1974& --\\
4&	4.4191&	4.8218 &0.60852& 0.91187    &3.2022& 2.9909\\
5&	4.2885&	4.623 &0.54216&	0.73034     &3.1983& 3.1521\\
6&	4.2558&	4.5464 &0.53053& 0.68741	&3.193& 3.1762\\
7&	4.1762&	4.3673 &0.49907& 0.57756	&3.1733& 3.2045\\
8&	4.1498&	4.3252 &0.49078& 0.56093	&3.1662& 3.2023\\
9&	4.076&	4.2031 &0.47382& 0.52577	&3.1222& 3.1337\\
10&	4.0452&	4.1755 &0.45697& 0.49899	&3.1266& 3.1728\\
11&	3.9955&	4.1267 &0.45243& 0.48209	&3.065& 3.1574\\
12&	3.9779&	4.1024 &0.44967& 0.47511	&3.0617& 3.1496\\
13&	3.9643&	4.0429 &0.44432& 0.45673	&3.0676& 3.1249\\
14&	3.9528&	4.0239 &0.43877& 0.45198	&3.0687& 3.1173\\
15&	3.9652&	4.0742 &0.44389& 0.48028	&3.0846& 3.1281 \\ \hline
\end{tabular}

\end{center}
\end{table}

\begin{table}
\caption{Numerical values of $StD$-based and $CVaRD_\beta$-based Sharpe ratios for various constant values of risk aversion parameter $1\le a\le 15$ in $U$ and $a_0=a$ and $a_1=a-3$ in $W$ (see Figure \ref{fig:riskmeasures_const_a}). \label{tab:sharperatios}}
\begin{center}
\scriptsize
\begin{tabular}{l|ll||ll|ll}
$a$ & $SR$ & $SR_{CVaRD}$ & $a_0$& $a_1$ & $SR$ & $SR_{CVaRD}$ \\ \hline \hline
 1  & 5.2805  &  2.6251  & -  & - & -  & - \\
 2  & 5.6693  &  2.8556  & -  & - & -  & - \\
 3  & 7.0539  &  3.5005  & -  & - & -  & - \\
 4  & 7.2620  &  3.6314  & 4  & 1 & 5.2878  &  2.6336 \\
 5  & 7.9100  &  3.9337  & 5  & 2 & 6.3299  &  3.1430 \\
 6  & 8.0218  &  4.0043  & 6  & 3 & 6.6138  &  3.3181 \\
 7  & 8.3680  &  4.1641  & 7  & 4 & 7.5616  &  3.7558 \\
 8  & 8.4555  &  4.2190  & 8  & 5 & 7.7108  &  3.8518 \\
 9  & 8.6024  &  4.2734  & 9  & 6 & 7.9942  &  3.9303 \\
 10 &  8.8522 &   4.4037 & 10 & 7 & 8.3679  &  4.1643 \\
 11 &  8.8312 &   4.2939 & 11 & 8 & 8.5600  &  4.2574 \\
 12 &  8.8463 &   4.3417 & 12 & 9 & 8.6346  &  4.3056 \\
 13 &  8.9222 &   4.4210 & 13 & 10 & 8.8518 &   4.4040 \\
 14 &  9.0088 &   4.4710 & 14 & 11 & 8.9028 &   4.4385 \\
 15 &  8.9328 &   4.5028 & 15 & 12 & 8.4830 &   4.3063 \\
 \hline
\end{tabular}

\end{center}
\end{table}

\section{Conclusion}
We compared the results of stochastic dynamic portfolio optimization based on ma\-xi\-mi\-za\-tion of the terminal utility. For varying terminal utility functions we evaluated Conditional value-at-risk deviation-based Sharpe ratio measuring riskiness of a dynamic portfolio. We analyzed how the $CVaRD$-based Sharpe ratio depends on the shape of a chosen utility function. We showed that employing a decreasing absolute risk aversion utility function yields higher mean returns but, at the same time, higher values of risk measured by standard deviation or conditional value-at-risk when compared to results corresponding to the constant absolute risk aversion utility function. We furthermore compared the portfolio performance results by means of the  conditional value-at-risk deviation-based Sharpe ratio. We showed that the risk-adjusted performance of the associated portfolios is worse for lower risk aversion.

\section*{Acknowledgements}
We thank the referees for their valuable comments and suggestions which improved presentation of the results. The authors were supported by VEGA 1/0062/18 grant and DAAD-Ministry of Education of Slovak republic grant ENANEFA.


\end{document}